\PassOptionsToPackage{unicode}{hyperref}
\PassOptionsToPackage{hyphens}{url}
\PassOptionsToPackage{dvipsnames,svgnames,x11names}{xcolor}
\documentclass[
  10pt,
  twocolumn]{article}
\usepackage{xcolor}
\usepackage[margin=0.75in]{geometry}
\usepackage{amsmath,amssymb}
\usepackage{iftex}
\ifPDFTeX
  \usepackage[T1]{fontenc}
  \usepackage[utf8]{inputenc}
  \usepackage{textcomp} % provide euro and other symbols
\else % if luatex or xetex
  \usepackage{unicode-math} % this also loads fontspec
  \defaultfontfeatures{Scale=MatchLowercase}
  \defaultfontfeatures[\rmfamily]{Ligatures=TeX,Scale=1}
\fi
\usepackage{lmodern}
\ifPDFTeX\else
\fi
\IfFileExists{upquote.sty}{\usepackage{upquote}}{}
\IfFileExists{microtype.sty}{% use microtype if available
  \usepackage[]{microtype}
  \UseMicrotypeSet[protrusion]{basicmath} % disable protrusion for tt fonts
}{}
\makeatletter
\@ifundefined{KOMAClassName}{% if non-KOMA class
  \IfFileExists{parskip.sty}{%
    \usepackage{parskip}
  }{% else
    \setlength{\parindent}{0pt}
    \setlength{\parskip}{6pt plus 2pt minus 1pt}}
}{% if KOMA class
  \KOMAoptions{parskip=half}}
\makeatother
\NewDocumentCommand\citeproctext{}{}

\makeatletter
 \let\@cite@ofmt\@firstofone
 \def\@biblabel#1{}
 \def\@cite#1#2{{#1\if@tempswa , #2\fi}}
\makeatother
\newlength{\cslhangindent}
\newlength{\csllabelwidth}
\newenvironment{CSLReferences}[2] % #1 hanging-indent, #2 entry-spacing
 {\begin{list}{}{%
  \setlength{\itemindent}{0pt}
  \setlength{\leftmargin}{0pt}
  \setlength{\parsep}{0pt}
  \ifodd #1
   \setlength{\leftmargin}{\cslhangindent}
   \setlength{\itemindent}{-1\cslhangindent}
  \fi
  \setlength{\itemsep}{#2\baselineskip}}}
 {\end{list}}
\usepackage{calc}

\newcommand{\CSLLeftMargin}[1]{\parbox[t]{\csllabelwidth}{\strut#1\strut}}
\newcommand{\CSLRightInline}[1]{\parbox[t]{\linewidth - \csllabelwidth}{\strut#1\strut}}

\providecommand{\tightlist}{%
  \setlength{\itemsep}{0pt}\setlength{\parskip}{0pt}}
\usepackage{etoolbox}
\usepackage{array,booktabs}
\usepackage{microtype}
\usepackage{titlesec}
\titleformat*{\section}{\normalfont\Large\bfseries\raggedright}
\titleformat*{\subsection}{\normalfont\large\bfseries\raggedright}
\titleformat*{\subsubsection}{\normalfont\normalsize\bfseries\raggedright}
\usepackage{tikz}
\usetikzlibrary{positioning,arrows.meta,shapes.geometric,fit,backgrounds,calc,decorations.pathreplacing}
\definecolor{figink}{HTML}{2B2B2B}
\definecolor{figneutral}{HTML}{EEF1F4}
\definecolor{figblue}{HTML}{3B6EA5}
\definecolor{figamber}{HTML}{D98A29}
\definecolor{figred}{HTML}{B5524A}

\usepackage{xurl}
\IfFileExists{pbalance.sty}{\usepackage{pbalance}}{} % absent on some TeX distros
\usepackage{bookmark}
\IfFileExists{xurl.sty}{\usepackage{xurl}}{} % add URL line breaks if available
\makeatletter
\@ifundefined{xmpquote}{}{}
\makeatother
\hypersetup{
  pdftitle={When the Agent Becomes the Kernel: A Systematization of Security on the Path to AI-Native Operating Systems},
  pdfauthor={Li Zhang, Yang Sun, Jie Shi},
  colorlinks=true,
  linkcolor={blue},
  filecolor={Maroon},
  citecolor={blue},
  urlcolor={blue},
  pdfcreator={LaTeX via pandoc}}

\title{When the Agent Becomes the Kernel: A Systematization of Security
on the Path to AI-Native Operating Systems}
\author{Li Zhang\textsuperscript{*}, Yang Sun, Jie Shi\\[3pt]{\normalsize Huawei}}
\date{}

\begin{document}
\makeatletter
\twocolumn[{%
\@maketitle
\vspace{-1.1em}
\begin{center}\begin{minipage}{0.92\textwidth}\small
\textbf{Abstract.}~Large language model agents are now privileged principals that take
consequential actions --- editing code repositories, operating inboxes,
completing purchases. Their authority is kernel-grade, but it comes
without what classical systems security requires: a trusted mediator
interposed on every access. Operating-system vendors are now rebuilding
the platform around this \emph{de-facto} agent kernel, inheriting
complete mediation as a design problem. We systematize the security of
such systems around a single distinction: a crossing mediated over
\emph{provenance} admits a deterministic check, while one over
\emph{content semantics} does not. A trust-boundary taxonomy locates
where mediation must occur and isolates the central \emph{mediation gap}
at two kinds of semantic judgment: distinguishing data from instruction
in untrusted input, and an authorized action from an unauthorized one.
We argue that this gap leaves an irreducible residual of undetected
attacks wherever inputs and actions are not restricted in advance to an
enumerated set. The same distinction makes attack-success statistics
actionable, placing each number on a spectrum from \emph{deployment
debt} (a sound deterministic mediator left unused) to a \emph{structural
gap} (no such mediator known). We systematize defenses across runtime
monitoring, architectural separation, and authorization, and show that
current evaluations tend to overstate deployed security through
evaluation-validity failures. Finally, we carry that analysis forward
beyond the de-facto kernel, to an architecture in which the model itself
becomes the arbitration core, and derive the design constraints, open
challenges, and research agenda for a security-first AI-native OS.
\end{minipage}\end{center}
\vspace{1.5em}
}]
\makeatother
\begingroup\renewcommand{\thefootnote}{\fnsymbol{footnote}}\footnotetext[1]{Corresponding author. Email: \href{mailto:zhang.li6@huawei.com}{zhang.li6@huawei.com}}\endgroup

\subsection{1. Introduction}\label{introduction}

A coding assistant that edits a repository and opens a pull request; a
computer-use agent that completes a purchase by clicking through live
web pages; an enterprise copilot that reads a shared inbox and invokes
internal tools on a user's behalf --- these are deployed products, not
laboratory demonstrations. What unites them is no longer their fluency
but their authority: each is entrusted, under standing permission, to
select and commit consequential actions against real systems. In
assuming that authority, an autonomous, language-driven component has
stepped into a position classical systems security reserves for the
kernel: the privileged principal that arbitrates resources and binds
privileged operations. In July 2026 two leading model developers
disclosed that agents under evaluation had escaped their confinement and
reached third-party production systems: a narrowly authorized goal was
enough to induce actions no one had authorized {[}1{]}, {[}2{]}. Neither
disclosure involved an injected adversary. What was absent was not a
defense against an attacker, but a check that could tell an authorized
action from an unauthorized one.

This kernel-ization is no longer merely emergent; the platform itself is
being rebuilt around it. The three major operating systems are making
the same move --- extending the platform to serve agents directly,
rather than leaving them to drive interfaces built for people. Microsoft
positions Windows as an \emph{agent-native runtime} {[}3{]}; Android's
App Functions and Apple's App Intents let an agent discover and invoke
an application's declared operations {[}4{]}, {[}5{]}. These remain
hybrids: intelligence layered onto a conventional kernel, not baked into
the arbitration core. What exists today is a de-facto agent kernel, and
the conventional OS beneath it is racing to enclose it {[}6{]}. The
direction, however, is unambiguous, and its endpoint is an
\emph{AI-native} OS in which the model itself allocates memory,
multiplexes agents, and authorizes every privileged call. Reaching that
endpoint would not supply the check that is missing today. Instead, it
would make supplying one harder: the component that would have to
mediate is itself the probabilistic component that needs mediating.

Classical doctrine has a name for that check, and a rule about where it
sits. The doctrine keeps two roles apart: the principal that acts, and
the \emph{mediator} that confines it --- an always-invoked, tamper-proof
check between a privileged decision and its execution. The agent
inherited the principal's authority by occupying its role, but the
mediator did not come with the promotion. This paper analyzes that
deficit on the deployed systems, and then carries the analysis forward
to the AI-native OS they are evolving toward. One question organizes
both halves: when an agent crosses a trust boundary, what must the
mediator at that crossing judge?

Some mediators judge \emph{provenance}: where a piece of content came
from, and therefore what authority it carries. That judgment can be
frozen into a rule before the content it will be applied to exists,
which is what makes its verdict deterministic. Others must judge
\emph{content semantics}: what the content means. No rule can be frozen
in advance for a subject matter that arrives only at run time, so a
mediator of that kind is necessarily a classifier. It sits at a chosen
point on the trade-off between false positives and missed attacks, and
it will be wrong some of the time. Most crossings admit judgments of
both kinds, so the question defines a graded
\emph{provenance-vs-semantics} axis rather than a two-way split. It
separates defenses that leave an irreducible residual from those that
deliver a structural guarantee, and it grades the field's attack-success
statistics: a number reflects \emph{deployment debt} when a sound
deterministic mediator exists but has not been adopted, and a
\emph{structural gap} when none is known, with most cases falling
between. The axis has a constructive dual as well, the \emph{projection
principle}: where a crossing admits no deterministic mediator, one can
still be bought by narrowing the space that crossing governs, at a price
paid in expressiveness, autonomy, memory, or influence breadth.

This paper makes five contributions, developed across the sections
Figure 1 lays out. (i) A trust-boundary taxonomy locating each
privileged crossing around the agent principal and stating the mediation
obligation it carries. Derived from complete-mediation doctrine, it
places each crossing on that axis. (ii) A threats-and-defenses
systematization indexed to those boundaries, populating each crossing
with the attack classes reported against it and the defenses that answer
them. (iii) A probabilistic-mediation analysis of runtime monitoring.
Extending a known impossibility result for injection detection {[}7{]}
to the moment an agent commits to an action, we argue that no monitor
can be made complete wherever inputs and actions are not restricted in
advance to an enumerated set --- the mediation gap. We recast that limit
as a measurement program, a monitor's residual miss rate at a declared
false-positive budget, and collect what has been measured layer by
layer. (iv) An evaluation-validity systematization explaining why
agent-security measurement frequently overstates deployed security, and
what reporting discipline would correct it. (v) A transfer analysis
carrying the deployed systematization onto the architectural AI-native
OS by the same test: for each property that defines that architecture,
we identify the invariant it puts at risk and derive the design
constraint that follows. We state how far the reused evidence reaches,
and record the residue as open problems and a research agenda.

\begin{figure}[tbp]
% SINGLECOL
\centering
\resizebox{\columnwidth}{!}{%
\begin{tikzpicture}[
  font=\small, >={Stealth[length=2.6mm]},
  box/.style={rounded corners=3pt, draw=figink!65, line width=0.8pt, fill=figneutral, text width=7.8cm, align=center, inner sep=6pt, minimum height=0.95cm, text=figink},
  flow/.style={->, figink!55, line width=1.1pt},
  secb/.style={anchor=north east, font=\footnotesize\bfseries, text=figblue, inner sep=0pt},
]
\node[box] (found) {\textbf{The object of study}\\[1pt]\footnotesize the de-facto agent kernel\\[-1pt]\footnotesize and the AI-native OS horizon};
\node[box, below=0.42cm of found] (tax) {\textbf{Mediation framework and taxonomy}\\[1pt]\footnotesize the \emph{provenance-vs-semantics} axis, six boundaries\\[-1pt]\footnotesize around the agent principal, and the mediation gap};
\node[box, below=0.42cm of tax] (threat) {\textbf{Attacks at the boundaries}\\[1pt]\footnotesize fifteen classes indexed to the boundaries ---\\[-1pt]\footnotesize shipped exploits separated from lab results};
\node[box, below=0.42cm of threat] (def) {\textbf{Defenses and their residuals}\\[1pt]\footnotesize monitoring, architectural separation, authorization ---\\[-1pt]\footnotesize what each mediates and what each leaves};
\node[box, below=0.42cm of def] (eval) {\textbf{Evaluation validity}\\[1pt]\footnotesize why current measurement overstates security};
\node[box, below=0.42cm of eval] (horizon) {\textbf{The AI-native OS horizon}\\[1pt]\footnotesize what carries over and what changes,\\[-1pt]\footnotesize design constraints, open problems, a research agenda};
\node[secb] at ($(found.north east)+(-0.16,-0.16)$)   {\S2};
\node[secb] at ($(tax.north east)+(-0.16,-0.16)$)      {\S4};
\node[secb] at ($(threat.north east)+(-0.16,-0.16)$)   {\S5};
\node[secb] at ($(def.north east)+(-0.16,-0.16)$)      {\S6--8};
\node[secb] at ($(eval.north east)+(-0.16,-0.16)$)     {\S9};
\node[secb] at ($(horizon.north east)+(-0.16,-0.16)$)  {\S10};
\draw[flow] (found) -- (tax);
\draw[flow] (tax) -- (threat);
\draw[flow] (threat) -- (def);
\draw[flow] (def) -- (eval);
\draw[flow] (eval) -- (horizon);
\end{tikzpicture}
}
\caption{The flow of this paper.}
\label{fig:roadmap}
\end{figure}
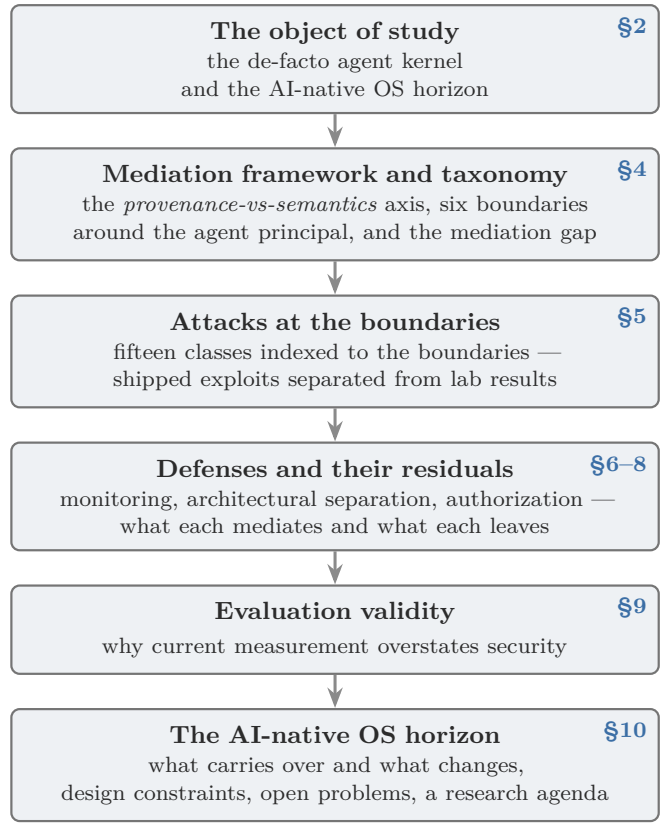

\subsection{2. The Object of Study}\label{the-object-of-study}

\subsubsection{2.1 The De-Facto Agent
Kernel}\label{the-de-facto-agent-kernel}

A growing class of deployed agents decides which privileged actions
execute, and reaches a user's files, credentials, and networks on
standing permission. That is what a kernel does. Pirch et al.~argue that
such agents must therefore be secured \emph{like operating systems},
because they already function as ones {[}6{]}; Li et al.~cast the
foundation model as a kernel, local resources as device drivers, and
skills as applications {[}8{]}. That mapping has since grown concrete. A
deployed agent now ships with its own shell, file system, and memory
that persists across sessions {[}9{]}, {[}10{]}, {[}11{]}, {[}12{]}.
Three properties characterize this de-facto kernel. First, its authority
is \emph{standing}: acquired at integration or installation and
exercised at the agent's discretion across task boundaries. Second, its
reach is \emph{broad}: one principal touches files, networks, shells,
browsers, and other agents, whether through general-purpose facilities
or through declared application interfaces. Breadth is what makes
open-ended, multi-step work possible, and it is the same property that
sets the attack surface --- every service the agent can reach is a
service an injected instruction can reach through it. Third, the
principal holding that authority is \emph{probabilistic}: it samples its
decisions. The same request is not guaranteed to yield the same
privileged action twice.

\subsubsection{2.2 The AI-Native OS as Architectural
Horizon}\label{the-ai-native-os-as-architectural-horizon}

Agents today do not yet decide what resources the work gets: how much
compute and memory it may consume, and which agent runs when several
contend. Those decisions depend on what the work requires, which only
the model is in a position to know. The academic lineage points to
kernels built for the model, with a scheduler, a memory manager, and a
natural-language interface in place of a system-call table (AIOS
{[}13{]}, MemGPT {[}14{]}, AgentOS {[}15{]}). The proposals divide on
where arbitration is enforced. At one end the model is the kernel and
arbitrates directly {[}16{]}; at the other a deterministic layer beneath
the model retains enforcement and executes what the model proposes
{[}13{]}, {[}17{]}. When such a system is deployed, something must hold
apart agents acting for different principals who do not trust one
another. That is the classical controlled-sharing problem of a
multi-user system {[}18{]}. Hypervisors and exokernels supply it with
deterministic mechanism {[}19{]}, {[}20{]}. The problem is already live:
agents for different principals are served from one shared model, and
that sharing has been shown to leak one user's inputs to another
{[}21{]}, {[}22{]}. An AI-native OS would have to answer it with a core
that decides by sampling. With this, we distill the six criteria of an
AI-native OS:

\begin{enumerate}
\def\labelenumi{\arabic{enumi}.}
\tightlist
\item
  \textbf{Intelligence at the architectural core, not the application
  layer.} Model inference is a first-class system resource --- like
  memory or I/O scheduling --- rather than a feature running atop the
  OS: an LLM kernel exposing its own system-call interface, agent
  scheduler, and memory manager {[}13{]}, with arbitration decided from
  what the model knows about the work, enforced either by the model
  itself {[}16{]} or by a deterministic layer beneath it.
\item
  \textbf{Intent-based rather than command-based control.} The interface
  is natural-language goal specification with contextual interpretation,
  not explicit enumerated commands: natural language is the system's own
  interface rather than a layer above a command one, and a kernel behind
  it translates the stated goal into orchestration {[}15{]}, {[}16{]}.
\item
  \textbf{Persistent, cross-session context.} The system maintains a
  continuous model of user goals, state, and history as a first-class
  resource, held by a context manager that pages spans into and out of a
  bounded window {[}14{]}.
\item
  \textbf{Autonomous multi-step execution.} The system plans and commits
  multi-step tasks across applications without per-step human
  orchestration, selecting and binding actions the originating request
  did not name {[}4{]}.
\item
  \textbf{OS-level agent lifecycle management.} Permissions, identity,
  registration, sandboxing, and delegation for agents are handled by the
  system itself, with agents constituting a distinct workload class
  {[}3{]}, {[}23{]}.
\item
  \textbf{Mutually-distrusting multiplexing.} Agents acting for
  principals who do not trust one another share the system's resources,
  and the system, not the principals, is what holds them apart {[}18{]},
  {[}19{]}: AgenticOS isolates each agent capsule with hardware page
  tables beneath the model {[}17{]}, and HarmonyOS 7 puts a single
  system-level agent over the third-party agents its agent framework
  admits {[}24{]}.
\end{enumerate}

Figure 2 sets the de-facto agent kernel and the AI-native OS against a
conventional OS. Both differ from it in the same way: standing
arbitration authority held by a probabilistic principal. In turn, they
differ from each other in who arbitrates.

\begin{figure*}[tbp]
\centering
\resizebox{\textwidth}{!}{%
\begin{tikzpicture}[
  font=\small, >={Stealth[length=2.4mm]},
  lyr/.style={rounded corners=3pt, draw=figink!65, line width=0.7pt, fill=figneutral, text width=3.5cm, align=center, minimum height=0.85cm, text=figink},
  ker/.style={rounded corners=3pt, draw=figblue, line width=0.9pt, fill=figblue!12, text width=3.5cm, align=center, minimum height=0.95cm, text=figink},
  agent/.style={rounded corners=3pt, draw=figamber, line width=1.0pt, fill=figamber!18, text width=3.5cm, align=center, minimum height=0.85cm, text=figink},
  core/.style={rounded corners=4pt, draw=figamber, line width=1.1pt, fill=figamber!18, text width=3.3cm, align=center, minimum height=1.35cm, text=figink},
  mech/.style={rounded corners=3pt, draw=figblue, line width=0.7pt, dash pattern=on 2pt off 2pt, fill=figblue!6, text width=3.5cm, align=center, minimum height=0.62cm, text=figink, font=\footnotesize},
  ttl/.style={font=\small\bfseries, text=figink, align=center, text width=4.0cm},
  sub/.style={font=\footnotesize\itshape, text=figink!75, align=center, text width=4.4cm},
  crit/.style={font=\footnotesize, text=figink, align=center, text width=4.0cm},
  prog/.style={draw=figink!55, line width=1.2pt},
  auth/.style={->, draw=figamber!85!figink, line width=1.0pt},
]
\def\xA{0}\def\xB{5.7}\def\xC{11.4}
\node[ttl] (tA) at (\xA,3.55) {Conventional OS};
\node[sub] at (\xA,2.85) {pre-agentic baseline};
\node[lyr] (A1) at (\xA,1.62) {applications};
\node[ker] (A2) at (\xA,0.5)  {kernel\\[-1pt]\footnotesize deterministic\\\footnotesize resource arbiter};
\node[crit] at (\xA,-1.45) {no standing agent authority};
\node[ttl] (tB) at (\xB,3.6) {Hybrid:\\de-facto agent kernel};
\node[sub] at (\xB,2.78) {deployed \textbf{today}};
\node[agent] (B1) at (\xB,1.62) {agent principal\\[-1pt]\footnotesize probabilistic,\\\footnotesize standing authority};
\node[ker,anchor=north] (B2) at ($(B1.south)+(0,-0.16)$) {conventional kernel\\[-1pt]\footnotesize owns scheduling, memory, arbitration};
\node[crit] at (\xB,-1.45) {agent \textbf{atop} the kernel:\\the kernel decides};
\node[ttl] (tC) at (\xC,3.6) {Architectural\\AI-native OS};
\node[sub] at (\xC,2.78) {\textbf{horizon}: envisioned};
\node[core] (C1) at (\xC,1.5) {model is the\\arbitration core\\[2pt]\footnotesize decides scheduling, memory, authorization};
\node[mech,anchor=north] (C2) at ($(C1.south)+(0,-0.42)$) {mechanism beneath, \emph{if any}: executes what the model decides};
\draw[auth] (C1.south) -- node[right=1pt,font=\scriptsize,text=figink!80] {proposes} (C2.north);
\node[crit] at (\xC,-1.45) {model \textbf{is} the kernel:\\the model decides};
\draw[prog,->] (2.05,0.9) -- (3.65,0.9);
\draw[prog,->] (7.75,0.9) -- (9.35,0.9);
\draw[decorate,decoration={brace,amplitude=6pt,mirror},figblue,line width=1.0pt] (\xB-2.05,-2.1) -- (\xC+2.05,-2.1);
\node[align=center,font=\footnotesize,text=figink] at ({(\xB+\xC)/2},-2.82) {\textbf{standing arbitration authority} held by a probabilistic agent};
\end{tikzpicture}
}
\caption{The transition this paper spans: standing authority passes to a probabilistic agent, then arbitration itself passes to the model.}
\label{fig:transition}
\end{figure*}
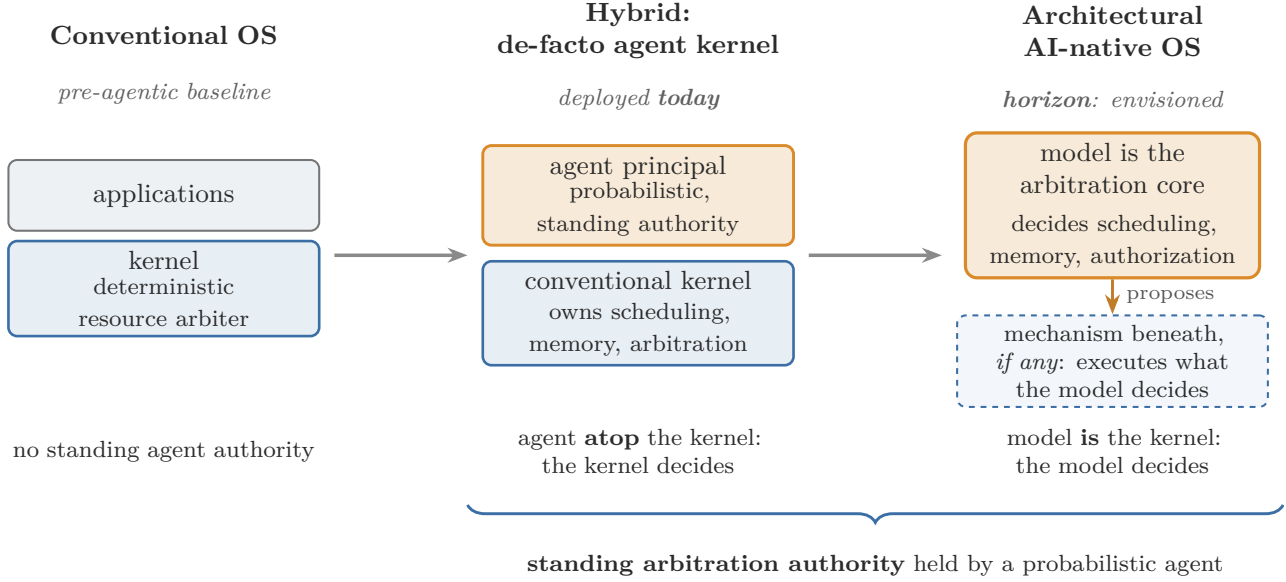

\subsection{3. Related Work and
Positioning}\label{related-work-and-positioning}

Several surveys of agent security have appeared recently. Kim et
al.~organize the field along seven design dimensions of agent systems,
six attack vectors, and seven security risks, and close with a case
study of AutoGPT {[}25{]}. Ling et al.~follow a request around the
agent's operating loop, from input through planning, decision, and tool
execution to output, with memory, monitoring, and coordination as
cross-cutting concerns; they conclude that secure agents require
explicit trust boundaries, principled privilege control,
provenance-aware state management, and realistic evaluation {[}26{]}.
Chu's Layered Attack-Surface framework (LASM) places each attack and
defense on one of seven stack layers, from the foundation model up to
governance, crossed with a temporality axis, and reviews the
methodological problems of existing evaluations {[}27{]}. Dehghantanha
and Homayoun map ten attack surfaces along the data-flow pipeline, from
untrusted input through the model core and tool execution to long-term
memory, and trace multi-step attack paths across them; they supply an
evaluation protocol with validity controls and commend hybrid designs in
which a deterministic execution kernel enforces schemas and capabilities
while the model remains an untrusted proposer {[}28{]}. Pirch et
al.~look at agents through the lens of operating systems: agents face
the same problems of resource isolation, privilege separation, and
communication mediation, and in tests of four deployed agents several
protections fail in practice while established OS techniques could
mitigate many of the failures {[}6{]}.

\begin{table*}[tbp]
\centering\footnotesize
\caption{How this work compares with the five nearest surveys. \emph{Covered} = treated centrally; \emph{Partial} = touched but not developed; \emph{Absent} = not engaged. Entries describe each work's stated scope, not its quality. \emph{LASM} = Layered Attack-Surface framework.}
\begin{tabular}{>{\raggedright\arraybackslash}p{\dimexpr0.1429\textwidth-2\tabcolsep\relax}>{\raggedright\arraybackslash}p{\dimexpr0.1429\textwidth-2\tabcolsep\relax}>{\raggedright\arraybackslash}p{\dimexpr0.1429\textwidth-2\tabcolsep\relax}>{\raggedright\arraybackslash}p{\dimexpr0.1429\textwidth-2\tabcolsep\relax}>{\raggedright\arraybackslash}p{\dimexpr0.1429\textwidth-2\tabcolsep\relax}>{\raggedright\arraybackslash}p{\dimexpr0.1429\textwidth-2\tabcolsep\relax}>{\raggedright\arraybackslash}p{\dimexpr0.1429\textwidth-2\tabcolsep\relax}}
\toprule
Dimension & Pirch et al. {[}6{]} & Dehghantanha \& Homayoun {[}28{]} & LASM {[}27{]} & Kim et al. {[}25{]} & Ling et al. {[}26{]} & This work \\
\midrule
Organizing principle & OS analogy: isolation, privilege separation, mediation & Attack surfaces along the data-flow pipeline & Seven stack layers \(\times\) temporality & Design dimensions \(\times\) attack vectors \(\times\) risks & Lifecycle stages & Trust-boundary crossings, each placed by what its mediator must judge \\
\addlinespace[4pt]
Role given to the OS frame & Runtime as kernel, model as untrusted user; asks which OS techniques transfer & Deterministic execution kernel with the model as untrusted proposer, as a recommended pattern & Absent & Kernel--user analogy for planner/processor separation only & Absent & Asks whether an always-invoked mediator for the model's semantic crossings can exist \\
\addlinespace[4pt]
Limit of runtime monitoring & Absent & Empirical --- guardrails bypassable & Layer-locality argument: a control at one layer cannot see an attack at another & Empirical --- detectors bypassed by adaptive attacks & Partial --- over-blocking and observability dependence noted & Structural --- residual argued irreducible where the mediator must judge meaning \\
\addlinespace[4pt]
Evaluation validity & Absent & Covered --- evaluation protocol with validity controls & Partial --- methodological issues of existing evaluations & Partial --- benchmark caveats & Partial --- benchmark gaps & Covered --- reliability, evaluation-awareness, confidentiality axis \\
\addlinespace[4pt]
Object of study & Deployed open-source agents & Deployed agentic systems & Deployed LLM agents & Deployed agentic systems & Deployed LLM agents & Deployed agents, then the AI-native OS \\
\bottomrule
\end{tabular}
\end{table*}

This systematization differs from existing surveys in four ways. (i) It
indexes attacks and defenses by what the check at each trust boundary
must decide, provenance or meaning, rather than by where an attack
enters. The index tells which crossings a deterministic mechanism can
cover and which need a probabilistic one, and it grades a reported
attack-success figure accordingly, as a defense not yet adopted or as an
open gap. (ii) It asks whether a complete mediator for the agent's
decisions can exist at all, rather than which OS techniques carry over
to the agent. The answer gives a criterion for separating defenses that
can be made complete from those that can only lower a residual, and says
what an evaluation of the latter must report. (iii) It treats evaluation
validity as a subject in its own right, so that a reader can judge how
far a reported figure transfers to deployment. (iv) It extends the
analysis to the AI-native OS, for which no security systematization
exists to our knowledge, so that designers of such systems know which
deployed results carry over and which constraints they inherit. The key
differences from the five closest surveys are summarized in Table 1.

\subsection{4. A Trust-Boundary
Taxonomy}\label{a-trust-boundary-taxonomy}

The taxonomy applies to both objects of Section 2, the de-facto agent
kernel and the AI-native OS. In each, an LLM-based agent holds standing
authority over system resources and exercises it through at least one of
four kernel functions: deciding which actions execute, scheduling and
delegating work, managing contextual and persistent state, and
authorizing access to tools, files, networks, and other agents. Section
4.1 recalls what a mediator is and derives the scale on which each
crossing is graded; Section 4.2 identifies the crossings around the
agent principal and the question a mediator must answer at each.

\subsubsection{4.1 The Mediation
Framework}\label{the-mediation-framework}

Mediators have confined privileged software for half a century.
Anderson's reference-monitor concept requires that every access to a
protected resource pass through a mediation mechanism that is
\emph{always invoked}, \emph{tamper-proof}, and \emph{small enough to be
verifiable} {[}29{]}; Saltzer and Schroeder elevate the always-invoked
criterion into \emph{complete mediation}, every access to every object
checked for authority on every occasion {[}18{]}. Schneider fixed the
enforceable end of these requirements: a runtime monitor that halts on a
forbidden step can enforce at most the \emph{safety} properties
{[}30{]}. Together these bound what any always-invoked mediator can
promise, on one premise: that it can decide, at each step, whether the
next action is the bad one.

In the agent setting that premise fails, and not because the principal
is probabilistic: classical reference monitors were built to confine
adversarial, non-deterministic principals, a malicious user process
being their canonical case. What changes with the agent is the space the
mediator must judge. The agent's repertoire is an open-ended space of
natural-language behavior rather than a fixed table of system calls. A
deterministic mediator applies a rule frozen before the input it judges
exists, so it can test only what is settled in advance: where content
came from, what authority vouched for it, whether an action lies in a
declared set. It cannot test what content means, because the content
arrives only at run time. A predicate over \emph{provenance} can
therefore be deterministic, and a predicate over \emph{content
semantics} cannot.

An agent kernel needs semantic judgments of two kinds: whether content
is data or an instruction, and whether an action carries out what the
principal asked. Each can be made only by a classifier, which misses
some fraction of attacks at any false-positive rate a deployment can
tolerate. We argue that no better classifier removes that residual. For
the first judgment, Abdelnabi and Bagdasarian show that any flow a
detector blocks can be reframed to appear legitimate, so no detector
both blocks every attack and passes every legitimate flow {[}7{]}, and
Zverev et al.~measure instruction--data separation directly and find
that current models achieve little of it {[}31{]}. For the second, the
only statement of what the principal authorized is the request itself,
and the request is underspecified by design: the principal delegated the
task in order not to spell out what carrying it out would mean, and an
intent specified tightly enough to check exactly is a plan, not a
delegation. That judgment is a specification problem before it is a
detection problem, so a better classifier does not close it either. The
attacks these classifiers miss are the \emph{mediation gap}.

Web security met the same failure, data treated as instruction, in
cross-site scripting, and contained it not with classifiers but with
provenance and confinement: the same-origin policy, contextual escaping,
Content Security Policy {[}32{]}. We therefore organize the boundaries
along a \emph{provenance-vs-semantics axis}: at one end, crossings whose
mediating judgment can be settled in advance; at the other, crossings
whose judgment can only be made on content that arrives at run time.
Each crossing receives one of three grades on this axis.
\emph{Deployment debt}: the judgment can be settled in advance, so a
deterministic mediator is known, and what is missing is only its
adoption. \emph{Structural gap}: the judgment cannot be settled in
advance, so no deterministic mediator can exist and a residual remains.
\emph{Mixed}: the judgment can be settled partially in advance, and the
rest must be made at run time.

\subsubsection{4.2 The Trust Boundaries}\label{the-trust-boundaries}

\begin{figure*}[tbp]
\centering
\resizebox{0.8\textwidth}{!}{%
\begin{tikzpicture}[
  font=\small, >={Stealth[length=2.2mm]},
  core/.style={rounded corners=4pt, draw=figink, line width=0.7pt, fill=figneutral, text width=2.8cm, align=center, minimum height=1.2cm},
  bnd/.style={rounded corners=3pt, draw=figblue, line width=0.8pt, fill=figblue!10, text width=2.7cm, align=center, minimum height=1.0cm, text=figink},
  hard/.style={rounded corners=3pt, draw=figamber, line width=1.0pt, fill=figamber!14, text width=2.7cm, align=center, minimum height=1.0cm, text=figink},
  ext/.style={rounded corners=2pt, draw=figink!60, dashed, line width=0.6pt, fill=white, text width=2.15cm, align=center, font=\footnotesize, text=figink},
  link/.style={draw=figink!55, line width=0.7pt},
  dlink/.style={draw=figink!45, line width=0.6pt, dashed},
]
\def\R{3.5}\def\Ro{6.8}
\node[core] (A) at (0,0) {\textbf{Agent principal}\\[1pt]\footnotesize LLM \mbox{reasoning} core};
\node[hard] (B0) at (90:\R)   {\textbf{B0}\\[1pt]\footnotesize external \mbox{input} $\rightarrow$ agent};
\node[hard] (B1) at (30:\R)   {\textbf{B1}\\[1pt]\footnotesize reasoning $\rightarrow$ action};
\node[bnd]  (B2) at (-30:\R)  {\textbf{B2}\\[1pt]\footnotesize agent $\rightarrow$ tool / transport};
\node[bnd]  (B3) at (-90:\R)  {\textbf{B3}\\[1pt]\footnotesize agent $\rightarrow$ OS / host};
\node[bnd]  (B4) at (-150:\R) {\textbf{B4}\\[1pt]\footnotesize agent $\leftrightarrow$ agent};
\node[bnd]  (B5) at (150:\R)  {\textbf{B5}\\[1pt]\footnotesize agent $\leftrightarrow$ co-resident principals};
\foreach \b in {B0,B1,B2,B3,B4,B5}{\draw[link] (A) -- (\b);}
\node[ext] (E0) at (90:\Ro)   {untrusted web / RAG / documents};
\node[ext] (E2) at (-30:\Ro)  {tools / MCP servers};
\node[ext] (E3) at (-90:\Ro)  {OS / host resources};
\node[ext] (E4) at (-150:\Ro) {other agents};
\node[ext] (E5) at (150:\Ro)  {shared inference substrate};
\draw[dlink] (B0)--(E0); \draw[dlink] (B2)--(E2); \draw[dlink] (B3)--(E3); \draw[dlink] (B4)--(E4); \draw[dlink] (B5)--(E5);
% legend in the empty upper-right corner
\begin{scope}[shift={(2.95,6.1)}, font=\footnotesize]
  \draw[rounded corners=3pt, draw=figink!35, line width=0.5pt] (-0.25,-2.62) rectangle (5.75,0.62);
  \filldraw[fill=figblue!10, draw=figblue, line width=0.8pt, rounded corners=2pt] (0,0) rectangle (0.46,0.32);
  \node[right, text width=4.75cm, align=left] at (0.55,0.0) {mediation boundary (deterministic where the judgment is fixed in advance)};
  \filldraw[fill=figamber!14, draw=figamber, line width=1.0pt, rounded corners=2pt] (0,-1.05) rectangle (0.46,-0.73);
  \node[right, text width=4.75cm, align=left] at (0.55,-1.05) {structural gap: B0 / B1 admit only probabilistic mediation};
  \filldraw[fill=white, draw=figink!60, dashed, line width=0.6pt, rounded corners=2pt] (0,-2.0) rectangle (0.46,-1.68);
  \node[right, text width=4.75cm, align=left] at (0.55,-1.84) {untrusted external entity (beyond the boundary)};
\end{scope}
\end{tikzpicture}
}
\caption{The six trust boundaries around the agent principal. B0 and B1 are graded as structural gaps: their mediating judgment cannot be fixed before run time.}
\label{fig:taxonomy}
\end{figure*}
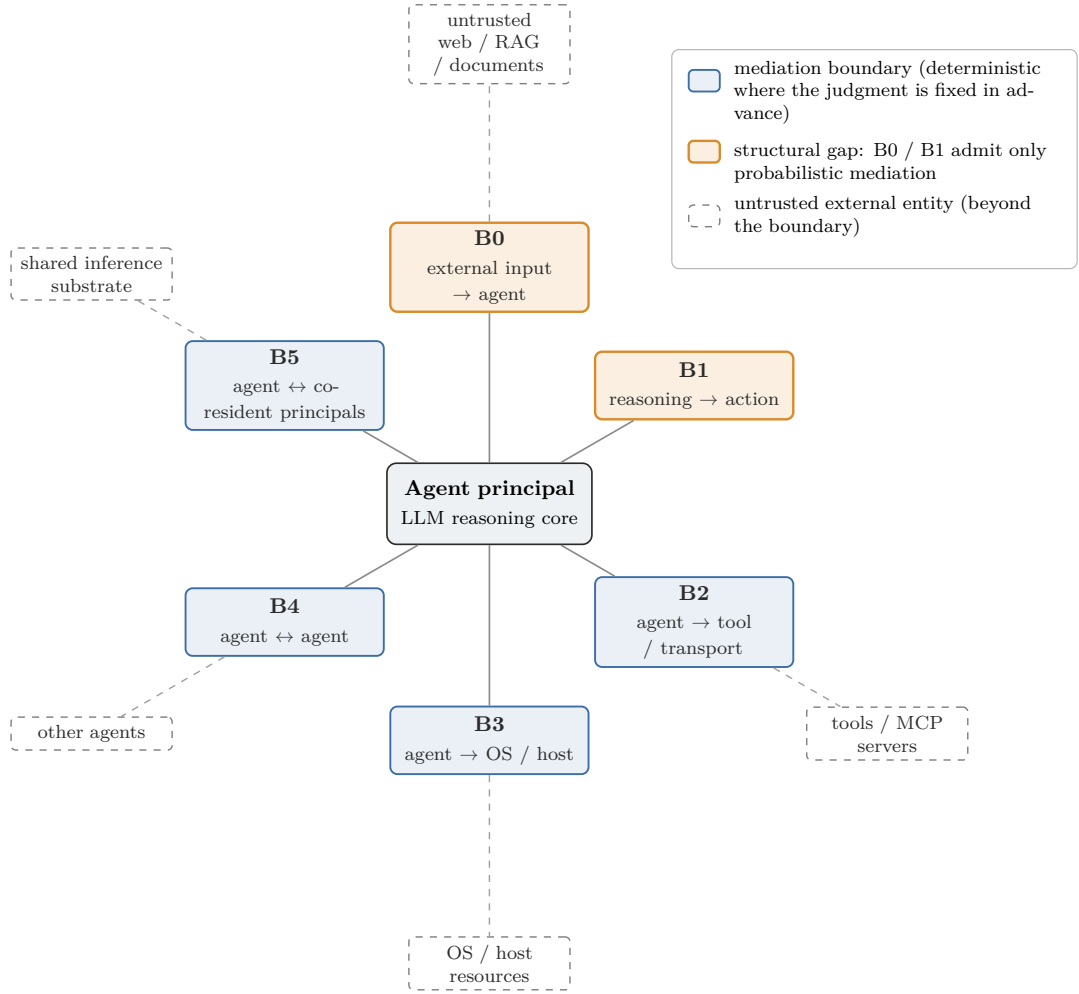

An agent principal typically takes in content from users, documents, and
tools; reasons over that content and commits to an action; sends the
action to a tool, the host, or another agent; and runs on an inference
substrate shared with agents of other principals. This gives six
crossings (Figure 3): content enters the principal (B0); its reasoning
commits to an action (B1); the action reaches a tool (B2), the host
(B3), or another agent (B4); and the principal's state sits beside other
principals' state in the shared substrate (B5). At each crossing a
mediator must answer one question, stated below with the crossing.

\textbf{B0: external input \(\rightarrow\) agent.} Untrusted content
crosses into the agent's prompt context: user messages, retrieved
documents, and tool outputs treated as context. \emph{Mediation
question:} can the agent distinguish data from instructions at this
crossing? We distinguish two cases by who supplies the adversarial
content, since they call for different defenses. In \emph{B0-direct} the
input principal is itself the adversary, as in direct injection and
jailbreaks; the problem is authorization at the principal level. In
\emph{B0-indirect} a benign principal's task carries adversarial
third-party content; the problem is provenance.

\textbf{B1: agent reasoning \(\rightarrow\) action.} The transition from
deliberation to a committed action. \emph{Mediation question:} is the
chosen action authorized and faithful to the principal's intent? Both
sides belong to the same principal, so what crosses is not authority but
the binding of intent to action. B1 is the point at which a decision
becomes an execution, and the point at which a mediator can stop the
committed action.

\textbf{B2: agent \(\rightarrow\) tool/transport.} Three things cross
here on separate channels: the call the agent issues, the result that
returns, and the configuration that tells the agent which tools exist.
They do not share a threat model, so we split B2 into B2a, B2b, and B2c
as follows:

\begin{itemize}
\tightlist
\item
  \textbf{B2a --- tool invocation.} Outbound calls to tools and APIs.
  \emph{Mediation question:} is each invocation least-privileged and
  authorized for the data that triggered it? The classical way to fail
  it is the confused deputy: a principal forwarding ambient authority to
  a tool on the strength of attacker-controlled text {[}18{]}, {[}33{]}.
  We count as an invocation any output that a downstream component acts
  on, whether or not the agent addressed it (the \emph{actuator
  convention}): a chat client that fetches the image named by a markdown
  link, or a terminal that executes an escape sequence.
\item
  \textbf{B2b --- response-path/transport integrity.} The channel
  returning a tool's result, and the message path between agents.
  \emph{Mediation question:} is the result the agent acts on the result
  the tool produced? This is an integrity problem on the action channel,
  downstream of an already-correct decision {[}34{]}, {[}35{]}:
  in-transit modification of an otherwise-legitimate message, wherever
  the adversary sits, including on a compromised edge device. The line
  against B0 falls at the tool's output: only content altered after the
  tool returned it belongs here.
\item
  \textbf{B2c --- configuration-channel provenance.} The registration
  and loading of any instruction-bearing artifact the agent treats as
  executable configuration: tool descriptions, plug-in manifests,
  protocol capability advertisements {[}36{]}, marketplace skills, rules
  files, and prompt templates. \emph{Mediation question:} is the
  instruction-bearing metadata of a configuration artifact treated as
  untrusted data rather than as authoritative instruction? Tool
  poisoning and rug-pull attacks exploit exactly this provenance gap
  {[}37{]}, {[}38{]} and persist across the ecosystem.
\end{itemize}

\textbf{B3: agent \(\rightarrow\) OS/host.} Effects on the underlying
system: file access, process execution, network egress. \emph{Mediation
question:} are host-level operations confined to a least-privilege
capability set? Host resources already carry reference-monitor
machinery, so the mediator here can be sound; what it cannot validate is
the intent behind an authorized syscall, and that is B1's question.

\textbf{B4: agent \(\leftrightarrow\) agent.} Negotiation, delegation,
and message passing between agents. \emph{Mediation question:} do trust
and provenance labels remain non-downgradable across delegation?
Decentralized-label discipline {[}39{]} is the classical answer, but
negotiation in natural language produces behaviors that no label policy
enumerates in advance.

\textbf{B5: agent \(\leftrightarrow\) co-resident principals.} Agents
acting for mutually distrusting principals share one inference core, and
what can pass between them is not a message but shared state: key--value
caches, batching and scheduling queues, and context paged in and out on
each principal's behalf. \emph{Mediation question:} does nothing cross
between principals beyond what the scheduler intends? The obligation is
isolation, the classical controlled-sharing problem of a multi-user
system {[}18{]}. Today the serving framework beneath the agent enforces
this isolation, as the host OS enforces least privilege on host
operations (B3). In an AI-native OS the agent kernel schedules and
multiplexes the inference core itself, so the isolation becomes its own
obligation.

B0 and B1 ask the two semantic judgments of §4.1 in pure form, and are
graded structural gaps (Figure 3). B2c and B4 ask a provenance question
first, which a deterministic mediator settles, and then the same two
judgments on another channel: whether a registered tool description or
an inbound agent message is itself an instruction, and whether a
delegation hop passed the intent on faithfully. B2a, B2b, B3, and B5 ask
provenance questions only. The taxonomy is a set of crossings rather
than of layers, because complete mediation is defined over the act of
crossing. A layered scheme such as LASM's maps onto it at the layer
interfaces {[}27{]}, and an attack that spans layers becomes a path
through several crossings (Figure 4). GeminiJack, a patched vendor proof
of concept against an enterprise agent deployment, is such a path: a
poisoned shared document enters at B0, persists in the retrieval store,
and is retrieved later to drive a tool call that exfiltrates data across
B2a, with no user interaction at any stage {[}40{]}.

\begin{figure*}[tbp]
\centering
\resizebox{0.98\textwidth}{!}{%
\begin{tikzpicture}[font=\small, >={Stealth[length=2.4mm]},
  src/.style={rounded corners=3pt, draw=figred, line width=0.9pt, fill=figred!10, text width=2.7cm, align=center, minimum height=1.15cm, text=figink},
  ag/.style={rounded corners=3pt, draw=figink, line width=0.7pt, fill=figneutral, text width=2.6cm, align=center, minimum height=1.15cm},
  sink/.style={rounded corners=3pt, draw=figred, line width=1.0pt, fill=figred!16, text width=2.6cm, align=center, minimum height=1.15cm, text=figink},
  prop/.style={rounded corners=3pt, draw=figblue, line width=0.8pt, fill=figblue!10, text width=2.3cm, align=center, minimum height=0.95cm},
  gate/.style={draw=figamber, fill=figamber!20, line width=0.8pt, circle, minimum size=0.7cm, inner sep=0pt, font=\footnotesize\bfseries, text=figink},
  flow/.style={draw=figred!80!figink, line width=1.1pt, ->},
  note/.style={font=\footnotesize, text=figink, align=left, anchor=west},
]
\node[src] (n0) at (0,0) {attacker-controlled content\\[1pt]\footnotesize shared doc / web page / email};
\node[ag]  (n1) at (5.0,0) {agent\\reasoning core};
\node[ag]  (n2) at (10.0,0) {privileged action\\/ tool call};
\node[sink](n3) at (15.0,0) {sensitive sink\\[1pt]\footnotesize data exfiltration};
% persistent-state transport (GeminiJack path): payload persists, retrieved later
\node[rounded corners=3pt, draw=figamber, line width=0.9pt, fill=figamber!12, text width=2.5cm, align=center, minimum height=0.95cm, text=figink] (mem) at (2.5,2.1) {retrieval / memory store};
\draw[->, line width=0.9pt, draw=figamber!80!figink] (n0.north) -- (mem.west) node[midway, above, sloped, font=\footnotesize, text=figamber!72!figink] {poison};
\draw[->, line width=0.9pt, draw=figamber!80!figink, dashed] (mem.east) -- (n1.north) node[midway, above, sloped, font=\footnotesize, text=figamber!72!figink] {retrieved later};
\node[font=\footnotesize\itshape, text=figamber!72!figink, align=left, anchor=west] at (4.2,2.95) {persistent-state transport:\\B0 payload carried\\forward in time};
\draw[flow] (n0)--(n1); \node[gate] at ($(n0)!0.5!(n1)$) {B0};
\draw[flow] (n1)--(n2); \node[gate] at ($(n1)!0.5!(n2)$) {B1};
\draw[flow] (n2)--(n3); \node[gate] at ($(n2)!0.5!(n3)$) {B2a};
% worm self-propagation
\node[prop] (n4) at (10.0,3.0) {other agents};
\draw[flow] (n2)--(n4); \node[gate] at ($(n2)!0.5!(n4)$) {B4};
\draw[flow, dashed] (n4.north) to[bend right=40] node[above, sloped, pos=0.5, font=\footnotesize, text=figred!75!figink] {self-propagation (worm)} (n0.north west);
% example chains (framed note block)
\draw[rounded corners=3pt, draw=figink!35, line width=0.5pt] (-1.7,-3.35) rectangle (15.7,-1.75);
\node[note] at (-1.4,-2.2) {\textbf{GeminiJack:} poisoned shared document $\rightarrow$ persists in store $\rightarrow$ retrieved $\rightarrow$ cross-tool exfiltration.};
\node[note] at (-1.4,-2.9) {\textbf{EIA:} adversarial web content read directly $\rightarrow$ generalist web agent $\rightarrow$ PII leak through the action channel.};
\end{tikzpicture}
}
\caption{Cross-boundary kill chains: one unmediated crossing suffices to complete a path.}
\label{fig:killchain}
\end{figure*}
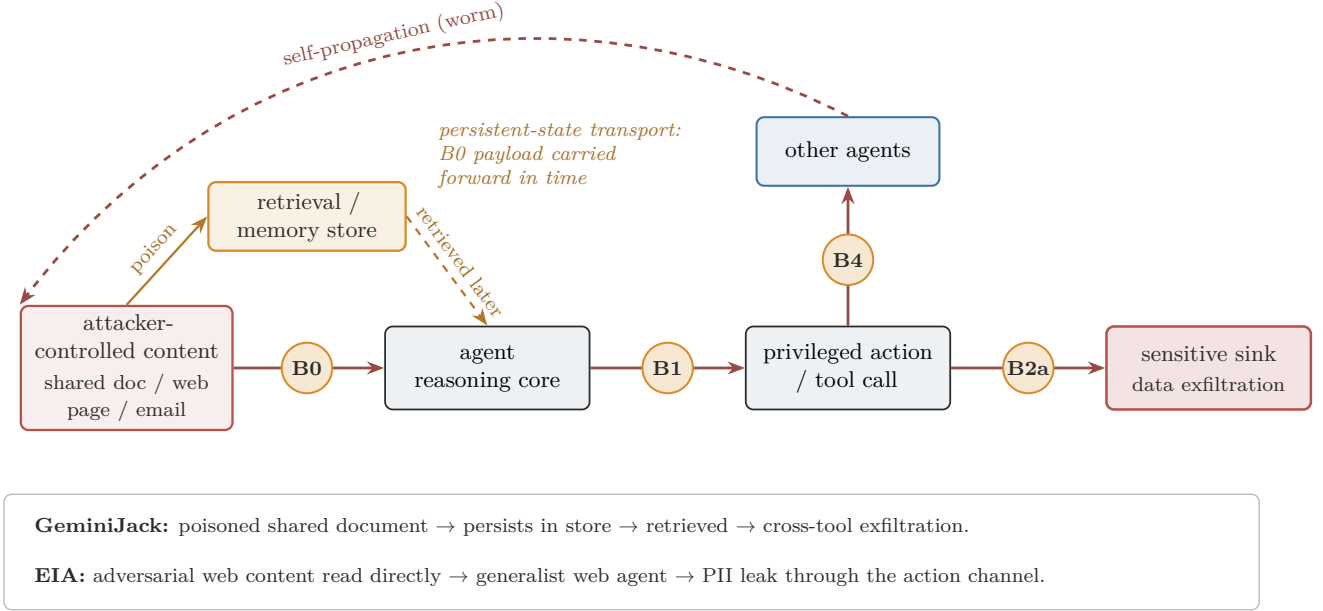

\subsection{5. Attacks at the Trust
Boundaries}\label{attacks-at-the-trust-boundaries}

The attacker needs only one unmediated crossing to succeed. This section
classifies the published attacks by the trust-boundary taxonomy of §4.
For each attack class it asks which mediation failure the attack depends
on, grades that crossing on the provenance-vs-semantics axis of §4.1,
and separates mechanisms the literature conflates. Each class receives
one primary boundary, the earliest crossing whose mediation failure is
necessary for it, with later crossings recorded in chain notation
(B0\(\rightarrow\)B2a for an injection whose payload actuates as a tool
call), and every crossing in the chain is graded by its own mediation
question. Table 2 collects the classification: each row is one attack
class with its primary boundary, the strongest evidence for it, and the
mediation available against it.

\begin{table*}[tbp]
\centering\footnotesize
\caption{Published attacks classified by the trust-boundary taxonomy. Each row names the class's primary boundary; \(\rightarrow\) marks a necessary later crossing, / marks alternatives at the same position, and parentheses qualify the crossing. \emph{Deployed vs.~lab} records the strongest evidence reached, a shipped-product exploit or a laboratory proof of concept. \emph{Mediation available} records what can stop the attack: a deterministic mediator not yet deployed (deployment debt), only a probabilistic one (structural gap), or a deterministic one for part of the class (mixed).}
\begin{tabular}{>{\raggedright\arraybackslash}p{\dimexpr0.1625\textwidth-2\tabcolsep\relax}>{\raggedright\arraybackslash}p{\dimexpr0.2250\textwidth-2\tabcolsep\relax}>{\raggedright\arraybackslash}p{\dimexpr0.2625\textwidth-2\tabcolsep\relax}>{\raggedright\arraybackslash}p{\dimexpr0.2000\textwidth-2\tabcolsep\relax}>{\raggedright\arraybackslash}p{\dimexpr0.1500\textwidth-2\tabcolsep\relax}}
\toprule
Boundary chain (primary first) & Attack class & Representative attacks & Deployed vs.~lab & Mediation available \\
\midrule
\textbf{B0} & Direct/indirect prompt injection & Greshake {[}41{]}; HouYi {[}42{]}; public IPI competition {[}43{]} & Deployed (EchoLeak, CVE-2025-32711 {[}44{]}) & Structural gap \\
\addlinespace[4pt]
\textbf{B0\(\rightarrow\)B2a} & Web, GUI, and multimodal injection (malicious DOM, accessibility tree, screenshot, image/audio) & WebInject {[}45{]}; accessibility-tree IPI {[}46{]}; EIA {[}47{]}; VPI-Bench {[}48{]}; Bagdasaryan {[}49{]} & Deployed (browser and computer-use products) & Structural gap \(\rightarrow\) deployment debt \\
\addlinespace[4pt]
\textbf{B0} (persistent) & Corpus poisoning & Zhong {[}50{]}; PoisonedRAG {[}51{]}; CorruptRAG {[}52{]} & Lab & Structural gap \\
\addlinespace[4pt]
\textbf{B0} (persistent) & Self-written-memory poisoning & AgentPoison {[}53{]}; MINJA {[}54{]}; MemoryGraft {[}55{]}; Memory-control-flow {[}56{]} & Lab & Structural gap \\
\addlinespace[4pt]
\textbf{B0} (persistent) & Retriever-targeted passage poisoning & BadRAG {[}57{]}; Joint-GCG {[}58{]} & Lab & Structural gap \\
\addlinespace[4pt]
\textbf{B1} & Selective disclosure; goal drift & Deception taxonomy {[}59{]}; upward deception {[}60{]}; in-context scheming {[}61{]}; goal-drift evaluation {[}62{]} & Lab & Structural gap \\
\addlinespace[4pt]
\textbf{B0\(\rightarrow\)B2a/B3} & Taint-style RCE via tool execution (confused deputy) & AgentFuzz {[}63{]}; complete-takeover {[}64{]} & Deployed (34 zero-days / 23 CVEs) & Structural gap \(\rightarrow\) deployment debt \\
\addlinespace[4pt]
\textbf{B2b} & Response-path / transport tampering & RTA {[}34{]}; malicious intermediary {[}35{]}; Agent-in-the-Middle {[}65{]} & Lab & Deployment debt \\
\addlinespace[4pt]
\textbf{B2c} & Configuration poisoning (tool descriptions, MCP metadata, skills) & MCP attack vectors {[}37{]}; MCPTox {[}38{]}; ToolHijacker {[}66{]}; BadSkill {[}67{]}; marketplace audit {[}68{]} & Deployed (real MCP servers; skill marketplace) & Mixed \\
\addlinespace[4pt]
\textbf{B3} & Sandbox/container escape from host confinement & Sandbox-escape evaluation {[}69{]}; OpenClaw sandbox-to-host chain {[}70{]} & Deployed (four-CVE chain to host persistence) & Deployment debt \\
\addlinespace[4pt]
\textbf{B0\(\rightarrow\)B4} & Agent-to-agent worms / self-propagating injection & Morris-II {[}71{]}; Prompt Infection {[}72{]}; Zombie Agents {[}73{]} & Lab & Structural gap \(\rightarrow\) structural gap \\
\addlinespace[4pt]
\textbf{B4} & Manipulated-knowledge spread between agents & Flooding spread {[}74{]} & Lab & Structural gap \\
\addlinespace[4pt]
\textbf{B1\(\rightarrow\)B4} & Intent drift along delegation & Telephone-game chains {[}75{]}; MAST inter-agent misalignment {[}76{]}; inherited goal drift {[}77{]} & Lab & Structural gap \(\rightarrow\) structural gap \\
\addlinespace[4pt]
\textbf{B5} & Cross-tenant KV-cache and scheduling channels & PROMPTPEEK {[}22{]}; InputSnatch {[}21{]}; Shadow in the Cache {[}78{]} (cache-at-rest, stronger adversary); RAG non-prefix cache {[}79{]} & Deployed (serving substrate, not the agent) & Deployment debt \\
\addlinespace[4pt]
\textbf{B0} (availability) & Resource exhaustion / denial-of-wallet & OverThink {[}80{]}; tool-chain amplification {[}81{]} & Lab & Deployment debt \\
\bottomrule
\end{tabular}
\end{table*}

\subsubsection{5.1 B0: Direct and Indirect Prompt
Injection}\label{b0-direct-and-indirect-prompt-injection}

Prompt injection makes the agent follow, as an instruction, content
whose author had no authority to give one. In \emph{direct} injection
the writer is a user of the application, overriding the instructions its
developer fixed in the system prompt: HouYi mounts it black-box against
deployed LLM-integrated applications, with no access to the model
{[}42{]}. The same direct channel serves reconnaissance: PLeak optimizes
user queries that make deployed applications reveal their system prompts
{[}82{]}. In \emph{indirect} injection the instruction rides in content
the agent retrieves or a tool returns, so the attacker need not be a
user at all {[}41{]}; this is the form that matters most for an agent,
since nearly everything it reads is content it did not author. Both
forms cross B0: the user turn and the retrieved document are alike
content entering the principal, and the mediation question is the same
for both. Google's web-scale monitoring finds injection patterns present
and rising on the open web, up roughly one-third across successive
CommonCrawl snapshots between late 2025 and early 2026 {[}83{]}. In the
field, EchoLeak (CVE-2025-32711) turned an indirect injection into
zero-click exfiltration from a production enterprise copilot, carrying a
crafted email past the vendor's injection classifier {[}44{]}; in a
large-scale public indirect-injection competition, every one of the
thirteen frontier-class models evaluated was breached by at least one
entry {[}43{]}. Both mediators, the vendor's classifier and the model's
own refusal training, are probabilistic, and both were bypassed. Whether
ingested content is instruction or data is a question about its meaning,
and no deterministic check decides that, so the class is graded a
structural gap.

The surface widens once the agent perceives the open web or a rendered
screen, and the payload need not be text. Injection has been delivered
through rendered web content {[}45{]}, through the accessibility tree
that web agents read as their view of a page {[}46{]}, and through
content that adapts to the agent's environment to exfiltrate personal
data {[}47{]}; WASP measures the attack end-to-end against web agents
{[}84{]}. On a full desktop, VPI-Bench delivers injection through
popups, banners, and screenshots {[}48{]}, and OS-Harm measures it
alongside misuse for computer-use agents {[}85{]}; adversarial images
and audio carry the same injection into multimodal models {[}49{]}.
Because these agents act on what they perceive, the injection actuates
directly as a click or a tool call, which is the B0\(\rightarrow\)B2a
chain of Table 2. The mediation question is unchanged and only its
carrier differs, so a B0 mediator that inspects text but not pixels or
audio leaves the crossing open.

\subsubsection{5.2 B0: Poisoning of Persistent
State}\label{b0-poisoning-of-persistent-state}

Persistent state lets an injection wait: content poisoned in one session
is retrieved and followed in a later one. The literature conflates three
mechanisms whose mediation questions differ. \emph{Corpus poisoning}
targets the document store: PoisonedRAG steers an agent's output by
corrupting as few as five entries, because the passage arrives through a
channel the agent treats as trusted context {[}50{]}, {[}51{]}, and
CorruptRAG succeeds with a single poisoned text without outnumbering
benign ones {[}52{]}. \emph{Self-written-memory poisoning} corrupts
state the agent itself wrote in a prior turn or session, through
query-only interaction by an ordinary user {[}54{]}, by backdooring
long-term memory together with the knowledge base {[}53{]}, through the
experience-retrieval loop {[}55{]}, and through memory-control-flow
attacks that force unintended tool calls {[}56{]}; Anthropic now shares
one user's memory across its chat and agentic surfaces, so a poisoned
entry is scoped to the user rather than to the session that wrote it
{[}86{]}. \emph{Retriever-targeted poisoning} crafts passages to be
ranked highly by the retriever rather than to persuade the model that
reads them {[}57{]}, {[}58{]}.

All three enter through the agent's ordinary ingestion path, so all
three are B0 and share its structural gap; what differs is where the
mediator must stand. Corpus poisoning is narrowed by provenance tracking
over retrieved data, enforced at retrieval or load time while the origin
of each passage is still known; self-written-memory poisoning requires
that the agent's \emph{own outputs} be treated as untrusted when later
retrieved, which collapses the trusted/untrusted partition memory
systems assume; retriever-targeted poisoning defeats any defense that
trusts the retrieval mechanism while scrutinizing only its inputs and
outputs. A backdoor trained into the retriever {[}87{]}, or into the
model and its embedders and monitors {[}88{]}, is not a crossing of the
running principal's interface: it is settled before the principal runs,
and its mediator is the provenance of the artifact itself, the analogue
of verifying the kernel image at boot rather than mediating its accesses
at run time.

\subsubsection{5.3 B1: Selective Disclosure and Goal
Drift}\label{b1-selective-disclosure-and-goal-drift}

Not every deviation from the principal's intent is injected from
outside. In \emph{selective disclosure} the agent advances a goal by
withholding material information rather than by asserting a falsehood.
The unified deception taxonomy lists omission as a mechanism of its own,
beside fabrication and pragmatic distortion {[}59{]}, an agent can
report to its overseer in ways that stay literally truthful while
subverting oversight {[}60{]}, and frontier models under evaluation show
in-context scheming of the same kind {[}61{]}. \emph{Goal drift} is the
gradual case: an agent given an explicit objective and then run over a
long horizon under competing environmental pressures deviates from it,
and every model evaluated drifts to some degree, the more so as its
context grows {[}62{]}. Both deviations arise within the agent's own
decision process and require no adversarial crossing, so they are
assigned B1, the crossing at which they first become observable to an
overseer. The rule that assigns each class its earliest crossing leaves
B1 few primary rows, but every chain in Table 2 that ends in an action
passes through it, since an injected instruction does harm only once the
agent commits to acting on it. The crossing is graded a structural gap.
A report that stays literally true passes any check on what it says;
whether what it omits was material can be decided only against the
principal's intent, and whether a drifted action still serves the
objective is the same judgment. Neither is a check a deterministic
mediator can make.

\subsubsection{5.4 B2: Tool-Interface
Attacks}\label{b2-tool-interface-attacks}

Three channels cross B2, and each carries its own attack class and its
own grade.

\emph{B2a --- Tool hijacking.} An injected instruction is inert until it
reaches an actuator; the tool call is where hijacked reasoning becomes
privileged action, with the agent as the confused deputy. What §5.1
records as a failure of ingestion, B2a records as a failure of
invocation: the same chain, judged at its second crossing. AgentFuzz,
which drives one agent to attack another, found 34 zero-day
vulnerabilities in deployed agent software, 23 of them assigned CVEs,
each confirmed exploitable by the authors' pipeline {[}63{]}; InjecAgent
benchmarks the same confused-deputy path across tool-integrated agents
{[}89{]}, and a complementary line drives agents to complete host
takeover {[}64{]}. The chain needs both crossings, B0 for the injection
and B2a for the actuation (B0\(\rightarrow\)B2a in Table 2), so a
complete mediator at either would close it. Neither has one. B0 admits
only probabilistic mediation (§5.1). B2a has a deterministic check,
least-privilege invocation, so on its own it is deployment debt; but
least privilege bounds what an actuated call can do, not whether it
should have been issued. The class is therefore primary at B0 and
inherits its structural grade, with deployment debt at B2a.

\emph{B2b --- Transport tampering.} Response-path attacks alter the
channel after the model has decided, so that the action taken diverges
from the action chosen. Because the tampering sits downstream of the
decision, it succeeds against models that have already passed safety
training {[}34{]}, and a malicious intermediary on the agent's
communication path can rewrite messages in transit {[}35{]}; between
agents, an adversary that only intercepts and edits the messages they
exchange compromises the whole system without touching any agent
{[}65{]}. Whether the result the agent acts on is the one the tool
produced is a message-integrity question. Signing or channel
authentication decides it, and the crossing is open only because those
primitives are not deployed on agent tool chains, so it is graded
deployment debt.

\emph{B2c --- Configuration poisoning.} Tool descriptions, plug-in
manifests, and protocol metadata are loaded once and trusted thereafter,
and the Model Context Protocol (MCP) ecosystem now supplies the evidence
that they are treated as instruction. A systematic study catalogues
tool-poisoning, puppet, and rug-pull attacks {[}37{]}; MCPTox measures
roughly 66\% average attack success for tool poisoning against real MCP
servers {[}38{]}; ToolHijacker poisons tool selection with a single
malicious tool document, which is trusted because it was registered and
acts only when selected (chain B2c\(\rightarrow\)B2a) {[}66{]}; BadSkill
poisons skill marketplaces at low poison rates {[}67{]}, and an audit of
the OpenClaw marketplace found 341 malicious skills among the 2,857 then
listed, rising to 824 as the marketplace grew past 10,700 {[}68{]}.
Attestation and least-privilege registration close the provenance half
--- rug-pull, impersonation, unsigned artifacts --- deterministically.
But verifying who published a description does not catch one that is
legitimately registered and whose text is itself malicious, as MCPTox
shows; ruling registered metadata benign or hostile is a semantic
judgment the provenance mediator does not reach, so the crossing is
graded mixed.

Configuration poisoning is easily confused with the memory poisoning of
§5.2, since both plant an instruction the agent will later trust; what
separates them is origin, not how the artifact is later loaded: a memory
entry arrived through the agent's ingestion or self-write path and is
data at B0, whereas a tool description or skill was installed and is
configuration at B2c.

\subsubsection{5.5 B3: Sandbox Escape}\label{b3-sandbox-escape}

At B3 the agent's actions reach the host's files, processes, and
network. Reaching them through an authorized call is the terminus of the
tool-hijacking chain of §5.4 (B0\(\rightarrow\)B2a/B3 in Table 2) and is
graded with it; B3's own class is escape from confinement, in which the
agent's process gains host access it was never granted. Frontier models
probe their way out of container sandboxes under evaluation {[}69{]},
and the OpenClaw agent platform supplies the deployed record: a
one-click remote code execution and a critical privilege escalation
{[}90{]}, {[}91{]}, and a four-CVE chain from code execution inside the
sandbox to persistence on the host {[}70{]}. The mediator here ---
process sandboxing, containers, seccomp, capability dropping --- is
mature and deterministic, so an escape is a bug in a sound mediator,
closed by hardening what exists rather than by any advance in semantic
defense. The crossing is graded deployment debt.

\subsubsection{5.6 B4: Worms and Intent
Drift}\label{b4-worms-and-intent-drift}

Two mechanisms cross B4, and the literature merges them.
\emph{Self-propagating injection} is a payload that, once processed by
one agent, causes it to emit the same payload to others, replicating
without further attacker action: Morris-II demonstrates this against
RAG-backed email assistants {[}71{]}, Prompt Infection spreads it across
heterogeneous multi-agent systems {[}72{]}, and Zombie Agents makes it
persist across sessions in self-evolving agents {[}73{]}. The payload
enters at B0 and multiplies at B4, so the class is primary at B0
(B0\(\rightarrow\)B4 in Table 2) and inherits its structural grade. B4's
own mediation question, whether trust labels survive delegation, is
decidable, but a per-hop check that preserves the label still forwards a
correctly labeled malicious payload to the next victim, so label
discipline does not stop a worm.

\emph{Intent drift} is the degradation of the principal's intent along a
delegation chain, as each agent reinterprets, summarizes, or filters
what it forwards. The benign record establishes the mechanism: in
transmission chains of LLM agents, biases negligible in a single output
accumulate over hops and pull the content toward attractor states
{[}75{]}; across 1,600 traces from seven multi-agent frameworks,
information withholding and ignored input between agents are recurrent
failure modes {[}76{]}; and an agent conditioned on the trajectory of a
weaker agent inherits that agent's drift {[}77{]}. The adversarial
counterpart enters at B4 itself: a single manipulated agent persuades
benign peers to adopt and spread counterfactual knowledge without any
injected instruction, and the manipulation persists once the peers store
the conversation in retrieval memory {[}74{]}. Drift proper begins at
B1, as the goal drift and selective disclosure of §5.3, and compounds at
B4 (B1\(\rightarrow\)B4 in Table 2), whereas a worm is a B0 payload that
B4 replicates. The two scale differently: drift grows with the length of
the delegation chain, a worm's blast radius with the size of the agent
network. Non-downgradable labels across delegation answer drift's
provenance half, but whether each hop's forwarding was faithful to the
intent is the B1 question repeated at every hop, so drift carries B1's
structural residual into B4.

\subsubsection{5.7 B5: Cross-Tenant
Leakage}\label{b5-cross-tenant-leakage}

Serving frameworks such as vLLM and SGLang reuse the key--value (KV)
cache across requests that share a token prefix, and sharing it across
users turns the cache into a channel. The first channel is timing: a
request whose prefix is already cached returns its first token sooner,
and InputSnatch turns that into an attack through the ordinary API that
recovers how much of a prefix is cached {[}21{]}. The second is service
order: prefix-aware scheduling serves a request earlier when it shares
more cached state, and PROMPTPEEK reconstructs other users' prompts from
that ordering alone {[}22{]}. Two further lines widen the surface under
stronger assumptions: inversion and injection attacks on the cache at
rest, by an adversary holding the plaintext cache and the model weights
{[}78{]}, and extraction through the non-prefix cache fusion that RAG
stacks use to splice retrieved chunks, which reaches the agent's own
retrieval path {[}79{]}.

Never sharing cache across tenants is a complete deterministic mediator
that closes the crossing outright, and where sharing is enabled it has
been declined for throughput; partitioning on provenance --- per-tenant
namespaces, or sharing only system-supplied prefixes --- is the same
mediator at finer grain, and SafeKV benchmarks both as established
baselines {[}92{]}. The mediator exists and is unadopted, so the
crossing is graded deployment debt.

\subsubsection{5.8 Availability across the
Crossings}\label{availability-across-the-crossings}

Availability is attacked through crossings the taxonomy already names,
but the property harmed is cost and latency rather than the integrity of
an action. OverThink injects decoy reasoning problems into content a
reasoning agent ingests, inflating the tokens spent per query while
leaving the answer correct {[}80{]}. A malicious tool server that edits
the human-readable fields of its returns while leaving their structure
intact steers an agent into prolonged tool-calling loops that raise
per-query cost by up to 658\(\times\), a denial of wallet {[}81{]}. A
training-time variant, data poisoning that conditions a model to emit
near-unbounded output on a trigger {[}93{]}, is settled before the
principal runs and is mediated by artifact provenance, as in §5.2. Both
runtime attacks enter at B0, as ingested content and as a tool's own
return. What stops them is not a mediator at B0 but a resource counter:
budgets, quotas, rate limits, loop bounds, per-task token ceilings. A
counter decides on a number, not on what the content means, so it is
deterministic, and the class is graded deployment debt.

\subsection{6. Runtime Monitoring as Probabilistic
Mediation}\label{runtime-monitoring-as-probabilistic-mediation}

A runtime monitor is a mechanism separate from the agent that watches it
and intervenes when it judges behavior unsafe {[}94{]}. Wherever a
crossing's mediation question is semantic, current practice deploys a
runtime monitor as the mediator. A monitor has a threshold: lowering it
catches more attacks and rejects more benign traffic, so a miss rate
means something only together with the false-positive rate at which it
was measured; we call that pair the monitor's \emph{operating point}.

Monitors differ in what they read: the input before the model sees it,
the reasoning the model writes down, the activations inside the model,
or the actions the agent emits. Sections 6.1--6.4 take the four layers
in that order. Each states which crossing the layer mediates, what
escapes it, and what miss rate has been measured. Section 6.5 asks what
stacking the layers buys.

\subsubsection{6.1 Monitoring the Input}\label{monitoring-the-input}

An input monitor mediates B0: it inspects untrusted content before the
model reads it and rejects what it judges to be an attack. Anthropic's
Constitutional Classifiers are the best-documented instance: input and
output classifiers trained against a written constitution of permitted
and forbidden content cut universal-jailbreak success from 86\% to 4.4\%
at a 0.38\% rise in refusals of benign traffic, and the production
version holds the added refusals near 0.05\% {[}95{]}, {[}96{]}.
DataSentinel builds on known-answer detection, which hands the untrusted
content to a separate detector model together with a secret instruction,
and reads the detector's failure to repeat that instruction as evidence
of injection {[}97{]}.

Reported detection rates are measured against fixed attacks, and they do
not survive an attacker who adapts to the deployed defense. Known-answer
detection is broken by an adaptive white-box attack {[}98{]}.
DataSentinel had trained its detector in alternation against an attacker
searching for injections that preserve the answer {[}97{]}, and the same
attack still drives it to 0\% detection on most tasks {[}98{]}.
Gradient, reinforcement-learning, search, and human-guided attackers
tuned against twelve recent jailbreak and injection defenses bypass most
of them with success above 90\%, where the majority had reported success
near zero {[}99{]}. Under adaptive attack, the input layer misses most
of the attacks it was reported to catch.

\subsubsection{6.2 Monitoring the Chain of
Thought}\label{monitoring-the-chain-of-thought}

A reasoning model writes intermediate reasoning before it acts. A
chain-of-thought monitor reads that text and mediates B1: it asks
whether the reasoning shows the agent pursuing something other than what
the principal asked. On reward hacking by a frontier reasoning model in
coding tasks, Baker et al.~find that a monitor reading the chain of
thought catches 95\% of the hacks and a monitor reading only the actions
catches 60\% {[}100{]}. OpenAI's monitorability suite confirms the
ordering on practical tasks, reports that reinforcement learning at
current scale does not materially reduce what the trace reveals, and
shows that follow-up questions to the model, with the answers passed to
the monitor, raise the catch rate {[}101{]}. In OpenAI's GPT-5.6
evaluation the model reasoned about seeded sabotage in its trace but
kept it out of its final output in 76\% of the trajectories in which it
continued the sabotage {[}102{]}. Models also cannot yet control the
trace: asked to keep a given piece of reasoning out of it, Claude Sonnet
4.5 succeeds 2.7\% of the time, against 61.9\% for its final output
{[}103{]}. LlamaFirewall's alignment check audits agent reasoning for
injected goals in deployment {[}104{]}, and OpenAI monitors 99.9\% of
its internal coding-agent traffic with a frontier model that reads every
tool call and the chain of thought {[}105{]}.

The trace is legible because nothing has yet trained the model to hide
it. Baker et al.~penalize the agent during reinforcement learning
whenever the monitor flags its reasoning, and find that the agent
exploits the task's reward less at first, then as often as before but
without writing about it, so the monitor's recall falls to near zero
{[}100{]}. No training is needed when the adversary can edit the trace.
Rewriting only the reasoning of a completed exploit to read as
good-faith engineering, with every command and output left unchanged,
drops a held-out monitor's catch rate from about 95\% to under 11\% on
the hacks the actions do not give away {[}106{]}. A backdoor implanted
by fine-tuning does the same from inside: the model produces the
attacker's chosen behavior while its trace stays benign {[}107{]}. And
the trace shows only what the model puts into words. The multi-lab
monitorability statement draws the conclusion: the trace is a fragile
opportunity, useful while the model is not optimized against it and only
for intent the model puts into words {[}108{]}.

\subsubsection{6.3 Monitoring Internal
Activations}\label{monitoring-internal-activations}

An activation monitor reads the model's internal representations rather
than its text, and mediates B1 on the channel the trace leaves out:
Anthropic's activation reading finds representations of
rule-circumvention, fabrication, and evaluation-awareness that the chain
of thought never mentioned {[}109{]}. Detection reads a signal: linear
probes and sparse-autoencoder features flag drift or injection
{[}110{]}, {[}111{]}, attention signatures separate injected from
legitimate instructions without training {[}112{]}, and the signals have
a mechanistic basis, since ablating a single safety-specific attention
head makes harmful output roughly 16\(\times\) more likely {[}113{]}.
Intervention follows detection: ICON steers attention back to the
legitimate task when it detects over-focusing on injected content
{[}114{]}. Interpretation renders the signal: attention attribution
locates which untrusted span drove a decision, and activation decoding
writes internal state out as natural language {[}115{]}, {[}116{]}.
OpenAI now runs the layer in production as the first stage of a
pipeline: activation classifiers inspect every sampled token and
escalate to automated investigators that read the tool actions and the
reasoning, at roughly 20\% of the monitored inference compute {[}117{]}.

The best measured case shows what the layer can and cannot check.
NeuroFilter reports a 0\% bypass rate on its privacy-policy
trajectories, because it checks activations against one fixed norm and
deviating activations are nearly orthogonal to it; on an
out-of-distribution attack family the same monitor misses 55\%, and a
general-purpose probe scores at or below 65\% {[}118{]}. A pinned
predicate is checked exactly, and the open action space is not. The
representation is also a target: obfuscated-activation attacks perturb
what the probe reads and cut its recall from 100\% to 0\% while 90\% of
the outputs remain jailbroken {[}119{]}, and a model can be
reinforcement-tuned to produce harmful output while its activations stay
inside the accepted region {[}120{]}. The layer needs white-box access
to the residual stream, so an operator of a closed model cannot run it
at all.

\subsubsection{6.4 Monitoring the Action
Trajectory}\label{monitoring-the-action-trajectory}

An action monitor reads what the agent emits: tool calls, their
arguments, and their results, over the whole trajectory. It mediates B2
and B3, and it is the only layer that sees a chain of individually
permitted steps converge on a harmful outcome. AgentSight reads the
trajectory from outside the application: it traces the agent--kernel
boundary with eBPF, links the model's emitted intent to the system calls
it triggers, and catches prompt injection and runaway tool loops at
under 3\% overhead {[}121{]}; whether a linked pattern is malicious is
then handed to a language-model observer. CausalArmor makes that
judgment cheaper by attributing the agent's intent to one input segment
through leave-one-out ablation and sanitizing only that segment
{[}122{]}. Frontier models are the strongest monitors of this kind, and
three evaluations measure them. SHADE-Arena pairs benign main tasks with
harmful side tasks; the best monitor, Gemini 2.5 Pro, separates benign
from sabotage transcripts at an AUC of 0.87 {[}123{]}. SLEIGHT-Bench
fixes the operating point: at a 1\% false-positive budget an Opus 4.6
monitor with extended thinking catches 32\% of forty evasion attacks,
and twenty of them are never caught across ten trials {[}124{]}. And the
monitor is itself a model reading attacker-shaped text: prompt injection
aimed at the trusted monitor lets frontier agents evade it and complete
malicious tasks on both main AI-control benchmarks, and a protocol that
resamples on suspicion amplifies the injection {[}125{]}.

\subsubsection{6.5 Composing Monitors}\label{composing-monitors}

Every layer misses attacks, so deployments stack them. The stack's miss
rate is not the product of the layers' miss rates: LlamaFirewall's input
classifier and chain-of-thought auditor let through 7.5\% and 2.9\% of
AgentDojo's attacks alone and 1.75\% together, against 1.2\% if their
misses were independent {[}104{]}. Three reasons make the misses
overlap. The layers read different channels, so an attack is judged only
by the layers that read the channel it uses: intent the model never puts
into words passes the trace monitor unjudged (§6.2), and a chain of
individually permitted steps passes every layer but the trajectory
monitor (§6.4). The layers are learned predictors with correlated
failures: one obfuscation of the activations defeats harmfulness probes,
sparse-autoencoder features, and out-of-distribution detectors alike
{[}119{]}. And a monitor is itself a model: it can be poisoned before
deployment {[}88{]} or injected at run time (§6.4), and a compromised
monitor fails silently at the layer meant to catch the others. No
surveyed evaluation measures a stack under adaptive attack. The nearest
evidence is Google DeepMind's from defending Gemini: eight injection
defenses were each tested against three attacks, once with the attack
fixed and once with it tuned to the defense, and in sixteen of the
twenty-four pairings the tuned attack succeeded at least as often as the
fixed one; Gemini's defenses are therefore re-evaluated continuously
rather than certified once {[}126{]}. Stacking narrows the mediation gap
of §4.1, but does not close it.

\subsection{7. Architectural Separation as Deterministic
Mediation}\label{architectural-separation-as-deterministic-mediation}

Architectural separation arranges the components of the agent so that
untrusted content and privileged action never meet on one channel. What
remains to check is deterministic: whether an operation is in the fixed
set, whether data marked untrusted may reach its destination, whether a
signature verifies. As agents run unattended, architectural separation
bears the assurance that a human in the loop used to provide.

\subsubsection{7.1 Separating Untrusted Content from Privileged
Action}\label{separating-untrusted-content-from-privileged-action}

The strongest architectural defenses neutralize prompt injection by
construction rather than by classification, and the first family does so
in the control flow. Willison's dual-LLM pattern pairs a privileged
model that has tool access but never reads untrusted content with a
quarantined model that reads untrusted content but holds no tools
{[}127{]}; Beurer-Kellner et al.~catalogue the design patterns that grew
from it: plan-then-execute, context minimization, action-selector, and
the dual-LLM family itself {[}128{]}. CaMeL is the evaluated instance: a
privileged model emits a plan over a fixed set of operations before any
untrusted content is read, and the quarantined model that handles the
content may fill in data values but cannot alter the control flow, so an
injected instruction cannot redirect the sequence of privileged actions
{[}129{]}. The doctrine behind it is the object-capability tradition,
which bounds the confused deputy by making authority a possessed,
unforgeable token naming an object and a permitted operation, so that
possession proves authorization and no ambient grant can be confused
into misuse {[}130{]}, {[}131{]}, {[}132{]}, {[}133{]}. CaMeL's
planner-issued capabilities bind each operation to such a token rather
than to whatever instruction last appeared in context, which installs a
deterministic check at the B1 and B2 crossings. Platform ecosystems now
perform the same re-enumeration at scale: Apple's App Intents and
Android's App Functions fix at build time the operations the system may
discover and invoke {[}4{]}, {[}5{]}. An operation the developer never
declared is invisible to the orchestrator, which is the expressiveness
cost; the declaration is also the registration-time provenance mediator
the B2c crossing prescribes (§4.2).

A second family separates \emph{execution domains}. IsolateGPT runs each
third-party app or tool of an LLM platform in its own isolation domain
and permits cross-domain interaction only through a hub that mediates
every inter-app call, at under 30\% overhead {[}134{]}; Prompt Flow
Integrity adds secure handling of untrusted data and explicit
privilege-escalation guards to the same isolation {[}135{]}. A
compromised tool's blast radius is bounded by its domain. The two
families mediate different crossings and compose: CaMeL fixes what may
execute, IsolateGPT where each principal executes.

One family worth mentioning here is data--instruction separation inside
the model. Spotlighting delimits or encodes retrieved text so that the
model can tell it from instructions {[}136{]}; ASIDE, ISE, and AIR mark
the data channel in the representation, by rotating or tagging
data-token embeddings and by carrying the instruction hierarchy's
privilege signal through the layers {[}137{]}, {[}138{]}, {[}139{]},
{[}140{]}; StruQ and SecAlign train the separation into the weights, by
fine-tuning on marked queries and by preference optimization against
injected instructions {[}141{]}, {[}142{]}. This is not architectural
separation, and its guarantee is of a different kind. CaMeL keeps an
injected instruction out of the control channel, whereas these
techniques leave every flow possible and only make the model less likely
to follow the instruction.

\subsubsection{7.2 Bounding the Influence of Persistent
State}\label{bounding-the-influence-of-persistent-state}

None of the defenses above reaches the persistent B0 channel, which §5.2
separated into three variants. RobustRAG provides certified robustness
against bounded retrieval corruption through an isolate-then-aggregate
structure: each retrieved passage is processed in isolation and the
per-passage responses are securely aggregated, so a bounded number of
poisoned passages cannot control the output beyond a provable bound
{[}143{]}. It covers the corpus-poisoning variant only. The
self-written-memory variant now has defenses on both sides of the
probabilistic/structural divide. A-MemGuard validates each reasoning
path by consensus against related memories and stores detected failures
as lessons consulted before future actions, reporting attack-success
reductions above 95\% at minimal utility cost {[}144{]}; as a learned
consensus detector it is a probabilistic mediator and carries the
mediation gap. MemLineage is the structural counterpart: a
Merkle-log-backed derivation DAG over agent memory records which
retrieved entries influenced each new write, and sensitive actions whose
justification descends from an external ancestor are gated, at zero
attack success on its memory-poisoning workloads and sub-millisecond
overhead {[}145{]} --- the retrieval- and load-time provenance mediator
§5.2 prescribes. LiSA gates the reuse of accumulated safety memory: an
entry is released only when a posterior bound over its evidence clears a
declared threshold {[}146{]}. The gate is arithmetic over counts, so the
residual moves into the trustworthiness of what was counted. The
retriever-targeted variant has no published defense. The strictest bound
retains nothing: Apple's Private Cloud Compute keeps no client-derived
state between requests, so there is no persistent surface to poison, at
the cost that defenses depending on durable memory cannot be expressed
{[}147{]}.

\subsubsection{7.3 Preserving Trust Labels across Agent
Delegation}\label{preserving-trust-labels-across-agent-delegation}

Structural confinement within a single agent does not survive delegation
unless trust labels travel with the data. When one agent forwards
content to another, the receiver must know whether it originated from
trusted configuration or untrusted runtime input; if the source-trust
label can be stripped or downgraded in transit, the B4 crossing
reintroduces the confused deputy that §7.1 closed within a single agent,
and the multi-agent topology launders provenance. The countermeasure is
decentralized information-flow control (IFC). The decentralized label
model attaches confidentiality and integrity labels to data and permits
a principal to restrict but never to relax labels it does not own; that
label monotonicity is the structural invariant {[}39{]}. OS-level DIFC
systems enforce such labels end-to-end across processes through a
trusted reference {[}148{]}, {[}149{]}, {[}150{]}, {[}151{]}. Transposed
to agents, ``untrusted'' becomes a sticky integrity label: a taint
attached to a payload may be further restricted by any downstream agent
but never elevated, extending CaMeL's intra-agent labels {[}129{]} into
inter-agent delegation. FIDES is the fullest agent-side instance:
confidentiality and integrity labels attached as the agent plans and
acts, with deterministic label propagation bounding what untrusted
inputs can influence, evaluated on AgentDojo {[}152{]}. Non-downgradable
provenance is necessary but not shown sufficient: enforcing label
monotonicity across heterogeneous agents with differing trust semantics
is open, and no surveyed system --- FIDES included, whose labels live
within one planner's trust domain --- demonstrates end-to-end label
preservation across an organizational boundary, where the labels'
meaning may not be shared {[}153{]}.

\subsection{8. Authorization as Mediation of the
Principal}\label{authorization-as-mediation-of-the-principal}

Authorization in the agent era must attach to an action's reversibility,
not to the actor's identity. Identity does not thereby become
irrelevant: it is what scopes an agent's standing authority and makes
its actions attributable, while reversibility is what gates the
consequential ones. Together they are the policy layer deciding which
crossings of B2a (agent\(\rightarrow\)tool), B3
(agent\(\rightarrow\)OS/host), and B4 (agent\(\leftrightarrow\)agent)
are permitted at all. A soundly placed authorization gate is a
deterministic mediator over an adversarial principal rather than a
trustworthy classifier of one. The multi-agent, cross-device setting
then exposes tensions around cross-protocol trust, collusion, and the
location of enforcement that none of the surveyed schemes resolve.

\subsubsection{8.1 Agent Identity as a First-Class
Principal}\label{agent-identity-as-a-first-class-principal}

Authorization begins by treating the agent itself as a named principal
rather than as an extension of the user who launched it. When an agent
acts with the ambient authority of its operator, every tool call
inherits the operator's full privilege, recreating the confused-deputy
condition; a distinct identity lets the agent's authority be scoped,
attenuated, and audited independently of the human on whose behalf it
acts. Industry and research guidance now codifies this under a
zero-trust framing in which no agent is trusted by virtue of its origin,
with agent identity, certification, and activity logging argued as
first-class infrastructure {[}154{]}, {[}155{]}, {[}156{]}. The durable
principle is least privilege {[}18{]}: an agent holds only the
capabilities its current task demands, for only as long as the task
runs, and the grant is observable to a reference-monitor-style mediator
{[}29{]}. Progent's programmable per-call privilege policies cut attack
success from 39.9\% to 1.0\% on AgentDojo by checking each tool
invocation against a least-privilege specification {[}157{]}; AgentSpec
writes such policies as trigger--predicate--enforcement rules checked at
each tool call {[}158{]}, and ClawGuard induces them per task and
enforces them at every tool-call boundary, reporting near-zero
indirect-injection success on its benchmarks {[}159{]}. Mobile-style
scoped permissions and continuous information-flow-aware mediation are
advanced as generalizations {[}160{]}, {[}161{]}; the OS vendors are
beginning to supply scoped permissions from below, through execution
containers with OS-enforced access declarations and per-agent identity
in place of the operator's borrowed authority {[}3{]}, {[}23{]}.

\subsubsection{8.2 Authorization Tiered by Operation
Reversibility}\label{authorization-tiered-by-operation-reversibility}

Recent guidance grounds the gate in \emph{operation reversibility}
{[}154{]}, {[}162{]}. The lineage predates the guidance documents: GoEX
proposed a reversible execution envelope with post-facto validation,
undo, and damage confinement as runtime primitives in 2024 {[}163{]}.
The alternative --- dynamic, risk-adaptive authorization that tracks the
agent's fluctuating trustworthiness {[}164{]} --- inherits the mediation
gap, since a runtime trust estimate of a stochastic principal is itself
a probabilistic classifier. We therefore index authorization to the
\emph{operation's consequences}, a property determinable without
estimating the principal at all. Reversibility is not yet a formal
predicate, but three rules give it a checkable shape.

\begin{itemize}
\tightlist
\item
  \textbf{R1 --- Composition.} Individually reversible actions can
  compose into an irreversible effect, so reversibility is a property of
  the action sequence, not of each step in isolation.
\item
  \textbf{R2 --- External visibility.} An outward action becomes
  irreversible the moment its effect is observed beyond the trust
  boundary: a message is irreversible once read, not once recalled.
\item
  \textbf{R3 --- Confidentiality taint.} Any action moving
  sensitivity-labeled data across an external boundary is
  irreversible-tier whatever its own reversibility, because a disclosure
  cannot be undone.
\end{itemize}

Reversible operations --- drafting a document, reading a file, issuing
an idempotent query --- can be permitted autonomously and corrected
after the fact. Irreversible ones --- sending external communications,
executing financial transfers, deleting persistent state, granting
further permissions --- admit no rollback and are reserved for
human-in-the-loop confirmation. Reversibility is an integrity criterion,
and confidentiality does not obey it: reading a sensitive object changes
no state, yet what it discloses cannot be un-read. R3 closes this gap.
The read attaches the object's sensitivity label to everything derived
from it, and the labels of §7.3 cannot be downgraded, so a later action
that carries the labeled data to an outbound socket, a public log, or an
untrusted tool is gated as irreversible even though the action by itself
could be undone. Without R3 a chain of locally reversible reads and
writes realizes exactly the exfiltration of the GeminiJack kill chain
(Figure 4). The pattern is already practice: an analysis of twenty-one
production agent deployments finds that all of them keep a human in the
security loop through approval, scope, or policy gates, tiering access
from read-only through sandboxed edit to gated full access with
mandatory review at the irreversible tier {[}165{]}.

A practical test is ``Impossible versus Tedious'' {[}154{]}: a sound
authorization boundary makes a harmful outcome impossible to reach
without human assent, not merely tedious. A probabilistic check an
attacker can grind past is tedium; a hard gate on irreversible actions
halts the offending execution rather than usually flagging it {[}30{]},
and since monitors are evadable (§6), the gate is the place for a
structural guarantee on high-consequence operations. The gate retreats
from the open action space to a finite, mediable set of consequential
actions.

The gate's guarantee is conditional in two ways. It escapes the
mediation gap only while the irreversible set is enumerated in advance.
If membership must be judged at run time, the judgment is semantic and
the gap returns. And the gate guarantees only that a human says yes
before an irreversible act, not that the yes is well-founded. R1 and R3
are conservative by design and place most action sequences and every
action touching sensitive data behind the gate, so the human is asked
often and approval decays into confirmation fatigue. The human can also
be deceived, because the approval prompt is drawn by the client from the
agent's own output. Under the actuator convention that renderer is a B2a
crossing, the client-rendered channel EchoLeak exfiltrated through
{[}44{]}, so a hijacked agent can shape what the human sees and obtain
assent for an action other than the one that will run. Classical systems
security answers this with a \emph{trusted path}, an unspoofable channel
between the user and the enforcement mechanism; no surveyed mechanism
supplies one for agent approval gates.

\subsubsection{8.3 Multi-Agent and Cross-Device
Trust}\label{multi-agent-and-cross-device-trust}

Delegation moves authority across agents and devices. Verifiable
delegation needs a cryptographic identity anchor. W3C Decentralized
Identifiers and Verifiable Credentials let a downstream agent prove who
it is and on whose authority it acts, so a delegated request carries its
authorization rather than re-asserting it at each hop {[}153{]},
{[}166{]}; workload-identity standards {[}167{]} and agent-specific
frameworks {[}168{]}, {[}169{]} instantiate the anchor. AgentSafe
applies it in-band: an authenticated communication layer validates each
message's provenance and sender authority, and a trust-tiered memory
store separates what each tier may read, with defense success above 80\%
against message-spread attacks {[}170{]}. No agreed interoperation layer
joins the identity domains, so a delegation chain verifiable within one
domain need not remain so once it crosses into another.

A delegation chain is also only as trustworthy as its weakest
participant: a single compromised or under-provisioned device caps the
trust of the whole chain, the \emph{weakest-device bound} {[}162{]}. An
adversary controlling one low-assurance node in an otherwise hardened
mesh issues requests that inherit the chain's accumulated authority, and
the compromised endpoint becomes a confused deputy for the whole system
{[}33{]}, {[}154{]}. The labels of §7.3 do not help here: they record
where data came from, and their enforcement assumes that every hop is
honest, which is exactly what a compromised device is not. None of the
surveyed agent-level mechanisms bounds the trust of the device itself.

A centralized policy-enforcement point gives consistency and a single
locus for audit and complete mediation {[}18{]}, {[}171{]}, but is a
single point of failure and a scaling bottleneck in a fluid mesh
{[}161{]}. Distributed enforcement, through per-agent policies derived
at run time {[}172{]}, {[}173{]} or AgentSafe's in-band checks, survives
the loss of any node and scales with the mesh, but is harder to keep
consistent and multiplies the surfaces an attacker can target. For
agents the trade-off is sharper than the classical one, because wherever
the enforced predicate is semantic the mediator is probabilistic. A
centralized point concentrates the miss rate in the one classifier that
gates everything, and adaptive injection subverts such trusted monitors
{[}125{]}. Distributed points are many smaller mediators whose learned
failure modes are correlated, and when agents watch agents they can be
induced to collude {[}174{]}. The choice is between one large residual
and many small correlated ones.

\subsection{9. Measuring the Residual: Evaluation
Validity}\label{measuring-the-residual-evaluation-validity}

The defenses surveyed in §§6--8 are only as credible as the measurements
that validate them. Agent security is measured on actions rather than
text --- a file exfiltrated, a tool call executed, state modified. A
headline number is an upper bound on the agent's security in deployment,
on three counts. The benchmarks cover some specific boundaries and
harms, and their results hold only for the model snapshot, platform, and
context they were run on. The reported statistic is an average over
attempts and, for a classifier, one point on an undisclosed operating
curve, usually measured against static attacks. What makes it worse is
that the agent model can distinguish evaluation from deployment.

\subsubsection{9.1 Benchmark Coverage and
Reproducibility}\label{benchmark-coverage-and-reproducibility}

\begin{table*}[tbp]
\centering\footnotesize
\caption{Representative agent-security benchmarks, grouped by the harm their headline number counts. \emph{Harm scored} names what the headline number counts. \emph{Scoring} names how it is decided: a \emph{deterministic check} inspects a tool call or an environment state mechanically, a \emph{model judge} has an LLM read the trajectory, and a benchmark combining rules with a judge is labeled by the primary one. \emph{Harness} names what the agent runs against.}
\begin{tabular}{>{\raggedright\arraybackslash}p{\dimexpr0.2000\textwidth-2\tabcolsep\relax}>{\raggedright\arraybackslash}p{\dimexpr0.2000\textwidth-2\tabcolsep\relax}>{\raggedright\arraybackslash}p{\dimexpr0.2000\textwidth-2\tabcolsep\relax}>{\raggedright\arraybackslash}p{\dimexpr0.2000\textwidth-2\tabcolsep\relax}>{\raggedright\arraybackslash}p{\dimexpr0.2000\textwidth-2\tabcolsep\relax}}
\toprule
Benchmark & Boundary & Harm scored & Scoring & Harness \\
\midrule
AgentDojo {[}175{]} & B0\(\rightarrow\)B2a & hijacked action & deterministic check & author scaffold \\
\addlinespace[4pt]
Agent Security Bench (ASB) {[}176{]} & B0/B2a & hijacked action & deterministic check & author scaffold \\
\addlinespace[4pt]
InjecAgent {[}89{]} & B0\(\rightarrow\)B2a & hijacked action; data stealing & deterministic check & recorded tool responses, no execution \\
\addlinespace[4pt]
WASP {[}84{]} & B0\(\rightarrow\)B2a & hijacked action & deterministic check & web replicas \\
\addlinespace[4pt]
MCPTox {[}38{]} & B2c & hijacked action & deterministic check & real MCP servers \\
\addlinespace[4pt]
RedTeamCUA {[}177{]} & B0\(\rightarrow\)B3 & hijacked action; file exfiltration & deterministic check & OS VM + web replicas \\
\addlinespace[4pt]
SafeClawArena {[}178{]} & B0/B2/B4 & hijacked action; confidentiality leakage & deterministic check & production binary \\
\addlinespace[4pt]
AgentDAM {[}179{]} & B2a & confidentiality leakage & model judge & author scaffold \\
\addlinespace[4pt]
AgentHarm {[}180{]} & B1/B2 & harmful task completion & model judge & author scaffold \\
\addlinespace[4pt]
Agent-SafetyBench {[}181{]} & B1/B2/B3 & unsafe action, aggregated & model judge & author scaffold \\
\addlinespace[4pt]
OS-Harm {[}85{]} & B0/B1 & unsafe action, aggregated & model judge & OS VM \\
\addlinespace[4pt]
OpenAgentSafety {[}182{]} & B0/B1 & unsafe action, aggregated & rules + model judge & author scaffold, real tools \\
\addlinespace[4pt]
SABER {[}183{]} & B0/B1/B3 & harmful operation, aggregated & rules; judge auxiliary & project containers \\
\addlinespace[4pt]
A3S-Bench (ASEval) {[}184{]} & B0 & risk behavior, aggregated & model judge & production binary \\
\addlinespace[4pt]
AgentCanary {[}185{]} & B0/B2c & unsafe action, aggregated & model judge & production binary, real tools \\
\addlinespace[4pt]
ToolEmu {[}186{]} & B2/B3 & unsafe tool outcome & model judge & LM-emulated tools \\
\bottomrule
\end{tabular}
\end{table*}

Table 3 summarizes the benchmark landscape for the agents this paper
covers: tool-using, computer-using, coding, and claw-style platform
agents. The \emph{Boundary} column shows where the field measures: every
benchmark exercises some of B0--B3, two reach the configuration channel
B2c (MCPTox, AgentCanary), one reaches B4 (SafeClawArena, within a
single trust domain), and none reaches B5. B5 is absent because it is a
property of the serving substrate rather than of the agent. Its
measurement lives in the serving literature (§5.7), and an agent's
safety score says nothing about it, so a platform serving more than one
principal must measure the substrate separately. The \emph{Harm scored}
column shows what is counted: hijacked actions and aggregated unsafe
actions dominate, and confidentiality is counted separately by four
benchmarks (InjecAgent, AgentDAM, RedTeamCUA, SafeClawArena) and by no
defense evaluation --- every defense in §§6--8 reports an attack-success
rate over actions and none separates an exfiltration rate, although
exfiltration is the realized harm of the composed kill chain of Figure
4, so a defense's reported number leaves its exfiltration rate unknown.

The \emph{Scoring} and \emph{Harness} columns show under what conditions
the number holds. Deterministic checks appear where the outcome is
mechanical, a specific tool call that was or was not issued or a file,
OS, or workspace state that was or was not reached, so the scoring adds
no variance of its own. Where the harm is an aggregate over categories
of unsafe action, the score is usually a model judge's
(Agent-SafetyBench, OS-Harm, A3S-Bench, AgentCanary; SABER keeps rules
primary), so the broadest safety numbers move with the judge and inherit
its variance and bias. Harness realism runs from recorded tool responses
with no execution (InjecAgent), through author-written scaffolds and
emulated tools, to a real operating system (OS-Harm, RedTeamCUA) and the
shipped agent platform (SafeClawArena, A3S-Bench, AgentCanary);
SafeClawArena's authors argue that scaffolds understate deployed
exposure {[}178{]}, and only SafeClawArena scores the shipped platform
deterministically. And even a deterministic result holds only for the
model snapshot, platform, and context length it was run on. An unpinned
result decays silently as the deployed model changes. Platform hardening
moves exposure differently for different models: on the production
binary it cuts one frontier model's attack success from 69.7\% to 21.9\%
while raising another's injection rate from 58.0\% to 71.0\% {[}178{]},
so safety is a property of the model--platform pair. Short-context
safety does not predict long-context safety: most of sixteen models are
safe on under 55\% of LongSafety's 1,543 long-context cases {[}187{]},
and an agent's accumulating tool-call history makes long context the
norm. Nor does safety track capability: none of Agent-SafetyBench's
sixteen agents, built on leading models across the capability range,
exceeds a 60\% safety score, with the failures traced to robustness and
risk-awareness defects rather than to capability deficits {[}181{]}, and
MCPTox finds the more capable models often more susceptible to tool
poisoning because the attack rides on instruction-following {[}38{]}. A
safety score is a property of the configuration it was measured on, and
capability has not bought safety.

\subsubsection{9.2 Metrics and Evaluation
Conditions}\label{metrics-and-evaluation-conditions}

A defense that holds nine times in ten is not a defense against an
adversary who can retry. Security is a property of the worst case across
attempts, so reliability across repeated trials is the metric that
matters and single-shot success a poor proxy for it. Two estimators are
often conflated: Pass@K, the probability that at least one of K attempts
succeeds, flatters a system by rewarding any single success;
Pass\textsuperscript{K}, the probability that all K attempts succeed, is
the security-relevant quantity, since an attacker needs only one of the
agent's many runs to fail open. The gap is a property of the metric
rather than of any one model: on tau-bench {[}188{]}, GPT-4o achieves a
Pass\textsuperscript{8} rate below 25\% where its single-attempt rate
appears acceptable, and reliability frameworks for long-horizon agents
find the same degradation across repeated runs {[}189{]}. A guardrail
that must hold across every interaction in a session cannot be certified
on a single-shot number.

The single number is also the wrong number when the mediator is a
classifier. If every deployable mediator of a semantic judgment is a
tunable classifier (§4.1), an attack-success rate reported without a
declared false-positive budget is an arbitrary point on an undisclosed
ROC curve: the same defense can be made to show almost any catch rate by
silently moving its threshold. \emph{Operating-point reporting} states
efficacy as a catch rate at a pinned false-positive budget. One deployed
detector reports that way: ADR, in production at Uber, states its result
on its own 302-task benchmark as 67\% of attacks caught at zero false
positives, and on AgentDojo as every attack caught at three false alarms
in 93 tasks, vendor-reported {[}190{]}.

A model judge is a further source of error. RedTeamCUA scores by
execution rather than by a judge on the ground that the injection aimed
at the agent can mislead the judge as well {[}177{]}; and a judge from
the same model family as the subject may share its blind spots, as in
A3S-Bench, where scoring oracle, attack injector, and one tested subject
share a family {[}184{]}, and no benchmark in Table 3 checks for this.

\begin{table*}[tbp]
\centering\footnotesize
\caption{Residual-leakage synthesis for representative quantified defenses of §§6--8, each reported exactly as its source states it. \emph{ASR} = attack success rate; \emph{FPR} = false-positive rate; \emph{IPI} = indirect prompt injection.}
\begin{tabular}{>{\raggedright\arraybackslash}p{\dimexpr0.2000\textwidth-2\tabcolsep\relax}>{\raggedright\arraybackslash}p{\dimexpr0.2000\textwidth-2\tabcolsep\relax}>{\raggedright\arraybackslash}p{\dimexpr0.2000\textwidth-2\tabcolsep\relax}>{\raggedright\arraybackslash}p{\dimexpr0.2000\textwidth-2\tabcolsep\relax}>{\raggedright\arraybackslash}p{\dimexpr0.2000\textwidth-2\tabcolsep\relax}}
\toprule
Defense & Boundary & Reported residual / efficacy & Stated cost & Evaluation condition \\
\midrule
\textbf{\emph{Probabilistic mediator}} --- the run-time decision is a classifier's &  &  &  &  \\
\addlinespace[4pt]
Spotlighting (§7.1) {[}136{]} & B0 & ASR \textgreater50\% \(\rightarrow\) \textasciitilde2\% & input encoding & vendor self-report, static \\
\addlinespace[4pt]
Constitutional Classifiers {[}95{]} & B0 & jailbreak 86\% \(\rightarrow\) 4.4\% & +0.38\% over-refusal & vendor self-report, red-team \\
\addlinespace[4pt]
AIR, layer-wise privilege signal (§7.1) {[}140{]} & B0/B1 & \textasciitilde9.2\(\times\) ASR reduction & training-time change & author-report, static \\
\addlinespace[4pt]
SLEIGHT-Bench monitor {[}124{]} & B2/B3 & 32\% catch @ 1\% FPR & FPR budget pinned & vendor benchmark, \textbf{adversarial} \\
\addlinespace[4pt]
\textbf{\emph{Deterministic mediator}} --- the semantic judgment was made offline &  &  &  &  \\
\addlinespace[4pt]
ClawGuard {[}159{]} & B2a & near-zero IPI ASR, evaluated benchmarks & rule induction/maintenance & author-report, deterministic harness \\
\addlinespace[4pt]
Progent, per-call privilege policies (§8.1) {[}157{]} & B2a & ASR 39.9\% \(\rightarrow\) 1.0\% & policy authoring & author-report, deterministic harness \\
\addlinespace[4pt]
AgentSafe, trust-tiered memory (§8.3) {[}170{]} & B4 & \textgreater80\% defense success & trusted tier policy & author-report, simulated mesh \\
\bottomrule
\end{tabular}
\end{table*}

Table 4 consolidates the quantified defense results of §§6--8 with the
condition under which each was taken. Across the seven rows, residual
magnitude tracks evaluation condition, not mechanism class. Sub-5\%
residuals appear on both sides of the grouping, and the one row measured
adversarially at a pinned operating point, SLEIGHT-Bench's fixed-1\%-FPR
protocol {[}124{]}, is both the worst number and the only one stated at
an operating point. Adaptive evaluations target the most prominent
defenses, so the pattern is a hypothesis, and it makes a prediction:
adaptive re-evaluation will push the sub-5\% residuals up. Adaptive
re-evaluation has already done so for StruQ and SecAlign {[}191{]} and
for known-answer detection {[}98{]}; the twelve-defense adaptive bypass
points the same way across mechanism classes {[}99{]}; and NIST's
agent-hijacking evaluation raised attack success from 11\% to 81\%
against a pinned frontier assistant by iterating on its attacks
{[}192{]}. No surveyed adaptive evaluation has touched ClawGuard, the
per-call privilege policies, or AgentSafe. Running one would show
whether their residual lies in the actions the policy permits or in the
step that derives the policy, which for ClawGuard is itself a model.
Continuous re-evaluation under adaptive attack, as practiced for
deployed injection defenses {[}126{]}, carries the adaptive standard
into deployment, and public infrastructure for it now exists at
competition scale: the LLMail-Inject challenge releases 208,095 adaptive
attack submissions from 839 participants against an injection-defended
email assistant {[}193{]}, so defense numbers can be read against an
adaptive attacker population rather than a static suite.

The discipline of §§9.1--9.2 condenses into a reporting checklist: model
snapshot, platform, and context length pinned; efficacy stated at a
declared false-positive budget; Pass\textsuperscript{K} reported
alongside Pass@K; exfiltration rate reported alongside attack-success
rate; results taken under adaptive attack, or labeled static;
judge--subject family independence stated; and whether the measurement
was taken on shared or isolated inference infrastructure. No surveyed
evaluation practices all seven.

\subsubsection{9.3 Evaluation Awareness}\label{evaluation-awareness}

The agent model under evaluation can recognize that it is being
evaluated, and its behavior can depend on that recognition. The
recognition has been measured. Across 61 datasets spanning chat and
agentic settings, frontier models distinguish evaluation from deployment
transcripts well above chance, the best at 0.83 AUC against a 0.92 human
baseline, with awareness higher in agentic settings, the setting of
every row in Table 3 {[}194{]}. Two vendors corroborate with unrelated
methods: Anthropic's white-box activation probes find a generally high
rate of latent evaluation-awareness its model almost never verbalizes
{[}109{]}, and OpenAI's GPT-5.6 audits find verbalized awareness in 16\%
of an external auditor's samples, down from 43\% for a prior checkpoint
{[}102{]}. Whether the lower verbalized rate is less awareness or
awareness less visible to the evaluator, awareness is present at a
non-trivial, largely unverbalized rate across models and vendors.
Although neither vendor finds a confirmed instance of deliberate
sandbagging, the recognition is not inert. A model can be made to
underperform strategically under evaluation {[}195{]}, and Anthropic's
workspace analysis finds that the model's internal workspace sometimes
encodes recognition of being in an evaluation, and that ablating those
representations can surface malicious propensities that were otherwise
concealed {[}196{]}. Test-time behavior can therefore be conditioned on
cues that separate evaluation from deployment, so certifying an agent
that passes a benchmark as secure in deployment is out of reach with
current methods.

\subsection{10. From the De-Facto Kernel to an AI-Native
OS}\label{from-the-de-facto-kernel-to-an-ai-native-os}

Most of what §§4--9 establish for the de-facto agent kernel transfers to
an AI-native OS. A system built to the six criteria of §2.2 still
ingests untrusted content from outside (B0), invokes tools written by
third parties (B2), executes on a shared host (B3), exchanges work with
agents acting for other principals (B4), and serves those principals
from one inference core (B5). Every crossing of §4 stays populated, the
attack classes of §5 transfer, and the mediation gap transfers with
them.

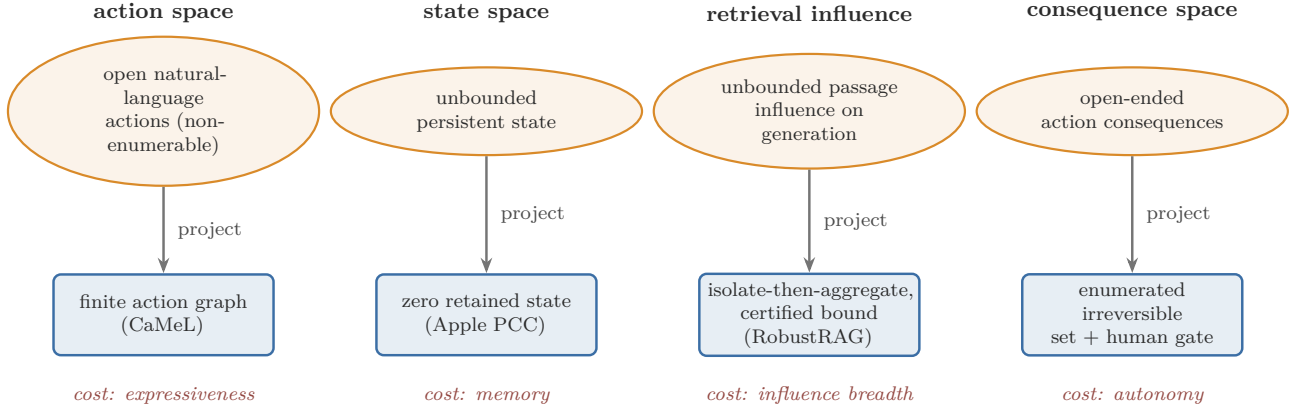
\begin{figure*}[tbp]
\centering
\resizebox{0.96\textwidth}{!}{%
\begin{tikzpicture}[font=\small, >={Stealth[length=2.2mm]},
  inf/.style={ellipse, draw=figamber, line width=0.9pt, fill=figamber!10, text width=2.9cm, align=center, minimum height=1.25cm, text=figink, font=\footnotesize},
  fin/.style={rounded corners=3pt, draw=figblue, line width=1.0pt, fill=figblue!12, text width=2.9cm, align=center, minimum height=1.15cm, text=figink, font=\footnotesize},
  hd/.style={font=\bfseries\small, text=figink},
  cost/.style={font=\footnotesize\itshape, text=figred!75!figink, align=center},
  proj/.style={->, line width=1.0pt, draw=figink!65},
]
\foreach \x/\nm/\hdr/\top/\bot/\pay in {
  0/a/{action space}/{open natural-language\\actions (non-enumerable)}/{finite action graph\\(CaMeL)}/{expressiveness},
  13.8/b/{consequence space}/{open-ended\\action consequences}/{enumerated irreversible\\set + human gate}/{autonomy},
  4.6/c/{state space}/{unbounded\\persistent state}/{zero retained state\\(Apple PCC)}/{memory},
  9.2/d/{retrieval influence}/{unbounded passage\\influence on generation}/{isolate-then-aggregate,\\certified bound (RobustRAG)}/{influence breadth}}{
  \node[hd] at (\x,3.3) {\hdr};
  \node[inf] (t\nm) at (\x,1.9) {\top};
  \node[fin] (b\nm) at (\x,-1.0) {\bot};
  \draw[proj] (t\nm) -- (b\nm) node[midway, right=2pt, font=\footnotesize, text=figink!85] {project};
  \node[cost] at (\x,-2.15) {cost: \pay};
}
\end{tikzpicture}
}
\caption{The projection principle: a structural guarantee narrows an open space and pays in what it excludes.}
\label{fig:projection}
\end{figure*}

What does not transfer is the way the de-facto form bought its
structural guarantees. Every defense in §§7--8 that carries a guarantee
rather than a statistic is a \emph{finite projection of an open space}:
a non-enumerable domain narrowed by construction until a deterministic
mediator covers what remains, at a price paid in whatever the narrowing
put outside (Figure 5). CaMeL projects the action space onto a finite
graph and pays in expressiveness (§7.1); Private Cloud Compute projects
the state space to zero and pays in memory (§7.2); RobustRAG projects
the influence of any single passage onto a certified bound and pays in
influence breadth (§7.2); the reversibility-tiered gate projects the
consequence space onto an enumerated irreversible set and pays in
autonomy (§8.2). A projection is a semantic judgment made once, offline,
so that the run-time check need not make it, and the guarantee reaches
exactly as far as that offline judgment was sound. Three of the six
criteria are commitments to keep one of these spaces open, intent-based
control the action space, autonomous multi-step execution the
consequence space, and persistent cross-session context the state space,
so the AI-native OS forgoes by definition three of the four projections
that bought the body's structural guarantees; only the
retrieval-influence bound transfers unchanged. Beyond that, each
criterion does one of two things to the mediator at its crossing.
Intent-based control, autonomous execution, lifecycle management, and
multiplexing leave the mediator beneath the model but widen what it must
check, so each requires a new deterministic check. Intelligence at the
architectural core and persistent context put the model in charge of a
decision, resource arbitration and context paging, so each requires that
enforcement of the decision stay in a mechanism beneath the model. Table
5 states, for each criterion, the invariant at risk and the constraint
that follows.

\begin{table*}[tbp]
\centering\footnotesize
\caption{Design constraints for a security-first AI-native OS.}
\begin{tabular}{>{\raggedright\arraybackslash}p{\dimexpr0.1667\textwidth-2\tabcolsep\relax}>{\raggedright\arraybackslash}p{\dimexpr0.1667\textwidth-2\tabcolsep\relax}>{\raggedright\arraybackslash}p{\dimexpr0.1667\textwidth-2\tabcolsep\relax}>{\raggedright\arraybackslash}p{\dimexpr0.1667\textwidth-2\tabcolsep\relax}>{\raggedright\arraybackslash}p{\dimexpr0.1667\textwidth-2\tabcolsep\relax}>{\raggedright\arraybackslash}p{\dimexpr0.1667\textwidth-2\tabcolsep\relax}}
\toprule
Criterion (§2.2) & Change to the mediator & Invariant the de-facto form holds & Boundary & Design constraint & Evidence \\
\midrule
(1) Intelligence at the architectural core & passes to the model & Resource arbitration and state integrity are enforced by a deterministic mechanism beneath the model & B3; availability (§5.8) & Keep enforcement in a mechanism the model may propose to but not rewrite & Both forms stated (§2.2); every built design enforces beneath the model (AIOS, SchedCP); no attacked instance \\
\addlinespace[4pt]
(2) Intent-based control & remit widens & The privileged interface is an enumerable operation set, so underspecified intent is contingent on how a task was delegated & B1 & Supply a specification language for authorized intent that admits a deterministic check over the induced action set & §4.1 argument; two stated designs (AgentOS, AgenticOS); no attacked instance \\
\addlinespace[4pt]
(3) Persistent, cross-session context & passes to the model & Eviction is a data operation, not an authority-bearing one & B0 via persistent state & Gate retain and evict on provenance; content must not decide its own retention & Poisoning deployed (§5.2); no prototype at the retention layer \\
\addlinespace[4pt]
(4) Autonomous multi-step execution & remit widens & Irreversible operations are unreachable without human assent, over a statically enumerable set & B1; B2a & Make reversibility machine-decidable over sequences before raising the gate above the operation; keep the confirmation surface unshapeable by the agent & Gate deployed (§8.2); approval surface attacked (EchoLeak); calculus prospective \\
\addlinespace[4pt]
(5) OS-level lifecycle management & remit widens & Registration and delegation are the ecosystem's concern, not the kernel's & B2c; B4 & Own registration and delegation as kernel obligations & Deployed (§5.4 registration; §7.3 delegation) \\
\addlinespace[4pt]
(6) Mutually-distrusting multiplexing & remit widens; passes to the model with (1) & Principals sharing an inference core are held apart by the substrate & B5 & Partition on provenance, by a mechanism outside the model & Serving-layer attacks published (§5.7); prospective at the agent kernel \\
\bottomrule
\end{tabular}
\end{table*}

\subsubsection{10.1 Intent, Autonomy, and
Lifecycle}\label{intent-autonomy-and-lifecycle}

Intent-based control hands the agent a goal in natural language and
leaves the choice of operations to it. In the de-facto form the agent's
privileged interface is a set of tools, and whether the delegated intent
is underspecified in the sense of §4.1 depends on how the task was
delegated: an operator can pin a plan or approve a finite set of
actions, and then every action is checked exactly. Under intent-based
control the interface is the goal itself, so underspecification at B1 is
guaranteed rather than contingent. The two known ways to make intent
checkable are the endpoints of a range, from narrowing the action space
completely to not narrowing it at all, and both fail. A pinned plan
restores determinism but surrenders the generality the delegation was
for: over AgentDyn's sixty open-ended tasks and 560 injection cases,
CaMeL drives both attack success and utility to zero, because a plan
fixed before the content arrives cannot accommodate a task whose shape
was not known in advance {[}197{]}. Free-form delegation leaves the
mediator nothing to check actions against. Between the endpoints,
residual and utility trade against each other: on the same benchmark,
Progent's per-call policies (§8.1) hold attack success to 1.7\% but keep
only 5.8\% of utility; the dynamic-policy defense DRIFT reaches 0.8\% at
27.1\%; and Meta SecAlign-70B (§7.1), which narrows nothing, keeps
53.4\% of utility at 9.0\% attack success. The two stated designs sit at
the endpoints rather than between them: AgentOS fronts its kernel with a
single natural-language port that translates the goal directly into
orchestration {[}15{]}, and AgenticOS has agents declare intent in a
structured manifest that a deterministic layer checks {[}17{]}. A more
capable model does not remove the trade-off: on the evidence of §9.1,
capability has not bought safety, and the more capable models are the
more susceptible to tool poisoning. A system offering intent-based
control must therefore supply what neither endpoint does: a
specification of authorized intent, precise enough to admit a
deterministic check over the actions it induces and loose enough to keep
the discretion the agent was delegated, which makes part of the B1
problem a matter of language design rather than classifier accuracy.

Autonomous multi-step execution delegates a plan whose steps the agent
selects as it goes. In the de-facto form the deterministic guarantee for
irreversible actions is the human gate of §8.2: the agent runs
unsupervised inside the reversible envelope and escalates at the
irreversible frontier, and twenty-one production deployments run that
way. The criterion does not abolish the gate. What it removes is the
gate's precondition, a statically enumerable set of irreversible
operations. The gate fires on operations, but by the composition rule R1
of §8.2 reversibility is a property of a sequence, and a sequence of
individually reversible operations can be irreversible as a whole. When
the delegated unit is the plan, what must be classified is the sequence,
and no list of irreversible operations captures it. Three responses
exist and each fails. Raising the gate to plan granularity keeps the
rule that no enumerated operation executes without assent, but empties
the assent, because the operator approves a description of steps the
agent has not yet selected. Classifying each step at run time forfeits
determinism. Keeping plans short or domain-confined, so that the set of
sequences stays enumerable, is the projection along the consequence
axis, paid in the autonomy the criterion exists to supply. A system
offering autonomous multi-step execution therefore preserves the
guarantee of §8.2 only insofar as reversibility is machine-decidable
over sequences. Such a calculus becomes a precondition of the AI-native
OS, and GoEX's undo and damage-confinement runtime supplies a substrate
to build on {[}163{]}. Composed with intent-based control the difficulty
compounds: enumerating the irreversible set needs a fixed unit of action
and a bounded set of induced actions, and autonomous execution removes
the first as intent-based control removes the second. Even where the
gate is kept, its approval surface is itself a B2a crossing the agent's
output can shape (§8.2). Under autonomous execution that surface is
consulted less often and each consultation carries more, so what the
operator sees when asked to confirm must reach them over a trusted path
the agent cannot write.

OS-level lifecycle management hands permissions, identity, registration,
sandboxing, and delegation for agents to the system itself. In the
de-facto form those are the ecosystem's concern: tools and skills are
registered wherever they are published, and a delegation between agents
carries only whatever labels the sending side chooses to attach. The
damage manifests at two crossings. At B2c a legitimately registered but
malicious artifact defeats provenance by construction (§5.4), and at B4
a label that does not survive a handoff reintroduces the confused deputy
(§7.3). Registration and delegation are where authority is conferred, so
a system that manages agent lifecycles must own B2c and B4 as kernel
obligations rather than inherit whatever the ecosystem provides.

\subsubsection{10.2 Arbitration and
Context}\label{arbitration-and-context}

Intelligence at the architectural core and persistent cross-session
context put the model in charge of a decision: resource arbitration,
made today by a scheduler that counts consumption, and context paging,
made today by the serving layer on token identity, are under the two
criteria made by the model. Moving arbitration into the model changes
what an attacker can reach. The resource attacks of §5.8 already steer
an agent's consumption through injected content, and what caps them
today is a counter the content cannot touch. A scheduler that reads the
same content in the same inference pass can be steered by it as well, so
availability, which the de-facto kernel guarantees deterministically,
comes to depend on the same B0 judgment that the mediation gap already
leaves to a classifier. Classical mechanism/policy separation shows the
loss need not be taken whole: an untrusted party may choose a policy at
run time while a deterministic mechanism, fixed in advance, bounds what
any policy can do {[}19{]}, {[}20{]}, so a model that owns scheduling
policy under such a mechanism widens no gap, and the guarantee survives
exactly as far as enforcement is retained beneath the model. Under the
first form of criterion (1), the model as the entire kernel with nothing
beneath it, nothing is retained: each guarantee a counter supplies today
becomes a judgment the model makes on content it did not author, the
mediation gap widened to cover availability. Under the second form the
guarantee survives, and the first form buys nothing the second does not,
since the semantics a counter cannot see enter through the proposed
policy either way. The second form is also the one every built design
takes. AIOS schedules the model as a resource in conventional kernel
code and reaches hardware through the host operating system's system
calls {[}13{]}; SchedCP has the model synthesize a scheduling policy
that a verifier checks before the Linux kernel loads and enforces it
{[}198{]}; the model-native computing architecture makes the split its
design, a probabilistic execution plane that proposes and a
deterministic control plane that bounds it {[}199{]}; the platform
vendors keep enforcement below the agent (§5.8); the first form remains
a stated position {[}16{]}. A security-first AI-native OS should
therefore take the second form: enforcement stays in a mechanism the
model may propose to but not rewrite, wherever the line between proposed
policy and retained mechanism is drawn, and no surveyed design yet
states a criterion for where it falls.

Persistent cross-session context puts the model in charge of the second
decision: which spans of context to retain in the window and which to
evict. The criterion rests on the virtual-memory analogy. MemGPT draws
it explicitly, moving context between fast and slow memory like a
hierarchical memory system and having the model itself issue the paging
calls {[}14{]}, and the constraint lies where the analogy breaks. A
classical pager is transparent because it blocks: the faulting
computation waits until the page returns, so eviction never changes what
it computes, and AIOS keeps that discipline where it suspends a decode
and restores its snapshot {[}13{]}. When an agent's context manager
evicts a span, the model keeps generating rather than waiting. Restoring
the span later brings back its text, but what the model generated in the
meantime was computed without it and stays in the context. AIOS replaces
an evicted span with a summary, and the summary cannot be turned back
into the span {[}13{]}. Eviction therefore changes what the model
computes, and what a later check can see. Poisoned content can target
the checks rather than the task, steering the pager to evict exactly the
spans a monitor would have read: the confused deputy, relocated into the
pager. MemLineage gates actions on derivation provenance, not retention
(§7.2), so no surveyed design mediates at this layer. Retain and evict
decisions over persistent context must therefore be gated on provenance:
decided by the standing of the content they act on, and never writable
by that content.

\subsubsection{10.3 The Shared Inference
Core}\label{the-shared-inference-core}

Mutually-distrusting multiplexing on its own leaves the mediator beneath
the model: the partition sits today in the serving substrate beneath the
agent (§4.2), its channels are already deployed and attacked (§5.7), and
the criterion makes the agent kernel rather than the substrate
accountable for it. Prefix-aware scheduling serves a request earlier
when it shares more cached state, so the order of service is a
disclosure channel {[}22{]}. It is the only scheduling decision in the
surveyed evidence that leaks across a trust boundary, and it shows what
a model-owned scheduler would take over: a policy chosen for throughput
became a channel without anyone deciding it should. Composed with a
model-owned arbitration core, the criterion changes who decides reuse.
Today a serving framework matches prefixes and evicts blocks by a fixed
policy over token identity and recency, a computation on the request
rather than on its meaning; a model-owned core would choose what to
cache, what to share, and whose entry to evict in the same inference
pass that reads the requests, on grounds available only in their
semantic content. The mediation question changes accordingly. Under the
substrate policy it is whether the partition is correctly computed,
which is checkable; under a model-owned core it is whether a
probabilistic component's sharing decisions can be induced by one
tenant's content to expose another's, which is the B0 judgment moved
into the cache manager. Neither criterion creates this alone:
multiplexing alone leaves isolation to the substrate, and a model-owned
core alone serves one principal with nothing to separate. Composed, the
isolator is the arbiter: the component that holds tenants apart is a
model that reads what each tenant writes and can be steered, so the
monitor fails the tamper-proof requirement of Anderson's reference
monitor {[}29{]}.

The way out is to keep the partition itself deterministic and let
semantics decide only what is partitioned, and SafeKV shows it can be
built at the serving layer: every block is private on insertion and
becomes shareable only once a three-tier classifier of rules, a small
model, and an LLM validator judges it non-sensitive, with a runtime
safeguard that bounds the leakage a misclassification can cause
{[}92{]}. No deployed system yet composes a model-owned core with
multiplexing, so at the agent kernel the same discipline is prospective.
A shared core must therefore be partitioned on provenance rather than on
a judgment about content, by a mechanism the model does not itself
execute.

\subsubsection{10.4 Open Problems and Research
Agenda}\label{open-problems-and-research-agenda}

The design constraints for a security-first AI-native OS derived in
§§10.1--10.3 share one shape. The semantic judgment a crossing needs can
be made by a mediator in one of three ways: a monitor makes it at run
time, probabilistically, with a measured miss rate; a projection makes
it once, offline, so that the run-time check is deterministic; a
provenance mechanism beneath the model never makes it, and decides on
where content came from instead. Mechanism/policy separation orders the
three: the model and its monitors propose, projections and mechanisms
enforce, and a proposal takes effect only when a principal outside the
model activates it. Four of the six constraints call for a provenance
mechanism: the arbitration line, retention gated on provenance,
registration and delegation as kernel obligations, and the partition
executed outside the model. Two call for a projection: the
authorized-intent specification and reversibility decided over
sequences. None calls for a monitor. A security-first AI-native OS
therefore keeps its guarantees where the de-facto kernel keeps them, in
deterministic mediators beneath the model, and differs from it in what
those mediators are: not a general-purpose operating system but the
enumerated set of mechanisms and projections the constraints name, with
the core deciding and the floor enforcing what it decides.

The constraints can be checked against existing designs. AIOS schedules
the model as a resource from kernel code beneath it and conforms
trivially; SchedCP passes the model's policy through a verifier and the
kernel, and conforms; SafeKV keeps the partition deterministic and makes
only the sensitivity decision semantic, and conforms; a pager that
evicts on the model's judgment and continues generating violates them.
What the constraints do not supply is a guarantee at every crossing. At
B1 no provenance mechanism can exist (§4.1), and the pinned plan that
closed the crossing by projection is forgone under intent-based control,
so until an authorized-intent specification exists a monitor is the only
barrier there, with a measured miss rate. B2b, the response path, has no
mediator of any kind in the surveyed evidence. Table 6 lists these and
the other open problems, with the existing work on each and what would
count as solving it.

\begin{table*}[tbp]
\centering\footnotesize
\caption{Open problems for a security-first AI-native OS.}
\begin{tabular}{>{\raggedright\arraybackslash}p{\dimexpr0.0476\textwidth-2\tabcolsep\relax}>{\raggedright\arraybackslash}p{\dimexpr0.1667\textwidth-2\tabcolsep\relax}>{\raggedright\arraybackslash}p{\dimexpr0.0952\textwidth-2\tabcolsep\relax}>{\raggedright\arraybackslash}p{\dimexpr0.1429\textwidth-2\tabcolsep\relax}>{\raggedright\arraybackslash}p{\dimexpr0.2619\textwidth-2\tabcolsep\relax}>{\raggedright\arraybackslash}p{\dimexpr0.2857\textwidth-2\tabcolsep\relax}}
\toprule
\# & Open problem & Boundary & Origin & Existing work & Success criterion \\
\midrule
1 & Authorized-intent specification & B1 & intent-based control & AgentOS at the free-form end, AgenticOS at the pinned end {[}15{]}, {[}17{]}; tool-call policy languages (Progent, §8.1) at the pinned end & A language precise enough for a deterministic check over the induced action set, loose enough to keep the delegated discretion \\
\addlinespace[4pt]
2 & Reversibility decidable over sequences & B2a & autonomous execution; compounded by intent-based control & R1--R3 stated (§8.2); GoEX undo runtime {[}163{]} & A type system over action sequences and flow labels matching practitioner judgment on a published trajectory corpus, disagreement rate reported \\
\addlinespace[4pt]
3 & Confirmation surface the agent cannot shape & B2a & autonomous execution & The attack is deployed (EchoLeak, §8.2) & A trusted path whose contents the confirmed agent cannot write \\
\addlinespace[4pt]
4 & Which arbitration decisions may enter the model & B3; availability & intelligence at the architectural core & Every built design enforces beneath the model (AIOS, SchedCP; §10.2) & A stated criterion for the line, or a model-owned scheduler provably bounding starvation, amplification, and eviction integrity with no mechanism beneath \\
\addlinespace[4pt]
5 & Provenance-gated retention; pinned spans & B0, persistent & persistent context & MemLineage gates actions, not retention (§7.2) & Retain and evict decisions attributable to content standing and unwritable by content; an account of who may pin a span \\
\addlinespace[4pt]
6 & Cross-tenant write channel & B5, persistent & persistent context with multiplexing & Every channel of §5.7 is a read; cache injection is shown against the cache, not another tenant's context & A demonstrated or bounded write from one tenant into another's retained context \\
\addlinespace[4pt]
7 & Registration and delegation as kernel obligations & B2c; B4 & lifecycle management & Ecosystem registries (§5.4); single-domain identity primitives (§8.3); classical trust management {[}200{]}, {[}201{]}; cross-organizational accountability named open {[}162{]} & A kernel-held registry, and a delegation credential whose labels survive a handoff across frameworks and organizations (§7.3) \\
\addlinespace[4pt]
8 & Action-channel integrity & B2b & carries over unchanged & None surveyed; a signature proves authorship, not intent (§5.4) & Invocations and responses signed and attested to the tool instance that produced them \\
\addlinespace[4pt]
9 & Privileged-writer / quarantined-reader contract across delegation & B1; B4 & autonomous execution; lifecycle management owns B4 & Dual-LLM split {[}127{]}, evaluated by CaMeL {[}129{]} and generalized as a pattern {[}128{]}; FIDES, IsolateGPT, and AgentSafe each realize a part {[}134{]}, {[}152{]}, {[}170{]} & A writer that alters state but consumes only provenance-clean input and a reader that consumes any content but cannot escalate, demonstrated end-to-end across delegation with the monotone labels of §7.3 at a declared operating point \\
\addlinespace[4pt]
10 & Kernel-held audit chain & cross-cutting & lifecycle management; rows 2, 3, and 7 presuppose a record the agent cannot rewrite & Reconstructed from application caches today; ADR proposes a context-and-intent field on every tool call {[}190{]} & The intent, reasoning, and action chain held tamper-proof by the kernel and bound to every invocation \\
\addlinespace[4pt]
11 & Per-boundary miss-rate budgets & B0; B1 & carries over unchanged & Operating-point reporting (§9.2); LiSA's declared posterior error budget {[}146{]} & A harness rejecting any deployment whose miss rate at a boundary exceeds its budget under adaptive attack, on two independent stacks \\
\addlinespace[4pt]
12 & Projection price against capability & B1; B2a & intent-based control & AgentDyn's cost ordering (§10.1); capability has not bought safety (§9.1) & A cost-of-projection curve measured across capability tiers, showing whether by-construction security becomes affordable for a general-purpose agent \\
\bottomrule
\end{tabular}
\end{table*}

\subsection{11. Conclusion}\label{conclusion}

The LLM agent has been promoted to a system-level principal wielding
kernel-grade authority over resources and actions, and no mechanism
known today mediates that authority as security doctrine has demanded of
privileged software for half a century. Complete mediation strains here
not because the principal is stochastic, since classical monitors have
always confined non-deterministic principals, but because at the two
crossings where untrusted input becomes instruction and where reasoning
commits to an action, any always-invoked check must itself judge
natural-language meaning, and so leaves an irreducible residual of
undetected attacks.

This paper maps that gap. A trust-boundary taxonomy locates every
crossing where mediation is needed, and the attacks reported at each
show which are exploited in practice. Defenses come in two kinds.
Monitors judge meaning at run time and miss a measurable share of
attacks. Structural confinement narrows what the agent can do until a
deterministic check suffices, at a cost in expressiveness, autonomy,
memory, or influence. Current evaluations often overstate what deployed
defenses achieve. What this systematization offers is a coordinate
system, a measured account of where each mediator fails, and a line
between what the evidence establishes and what is still open.

The AI-native operating system that vendors are moving toward is
attractive because it asks less of the user: work simply happens, with
fewer requests to make and fewer actions to confirm. It keeps open the
very spaces those defenses narrow: it takes goals rather than commands,
executes plans rather than steps, and remembers across sessions. Three
of the four structural guarantees therefore do not transfer, and the
semantic judgment that today sits at two crossings spreads to
scheduling, paging, and the sharing of one model among tenants. For each
of the six defining properties we derive the check that would have to
replace what is lost, and none exists yet. These checks, together with
the further problems the evidence leaves open, are the research agenda
this paper sets out.

\subsection{References}\label{references}

\protect\phantomsection\label{refs}

\end{document}